\documentclass[journal,letterpaper,onecolumn, 11pt, draftclsnofoot]{IEEEtran}
\usepackage{amsmath,epsfig,amsfonts,amssymb,mathrsfs,pifont,multirow,fmtcount}

\newtheorem{theorem}{Theorem}
\newtheorem{lemma}{Lemma}
\newtheorem{corollary}{Corollary}

\usepackage{cite}

\ifCLASSINFOpdf
\else
\fi
\usepackage{xspace}
\usepackage{bbm}
\usepackage{mathrsfs}

\newcommand{\Ac}{\mathcal{A}}
\newcommand{\Bc}{\mathcal{B}}

\newcommand{\Ec}{\mathcal{E}}

\newcommand{\Qc}{\mathcal{Q}}

\newcommand{\Tc}{\mathcal{T}}

\newcommand{\Xv}{{\bf X}}
\newcommand{\Av}{{\bf A}}
\newcommand{\av}{{\bf a}}

\newcommand{\Yv}{{\bf Y}}
\newcommand{\Zv}{{\bf Z}}
\newcommand{\Wv}{{\bf W}}

\newcommand{\Sv}{{\bf S}}

\newcommand{\wv}{{\bf w}}
\newcommand{\xv}{{\bf x}}
\newcommand{\yv}{{\bf y}}
\newcommand{\zv}{{\bf z}}

\newcommand{\bv}{{\bf b}}

\newcommand{\sh}{{\hat{s}}}

\def\e{\epsilon}

\DeclareMathOperator\E{\sf E}
\let\P\relax
\DeclareMathOperator\P{\sf P}

\def\textiid{i.i.d.\@\xspace}
\newcommand\iid{\ifmmode\text{ i.i.d. } \else \textiid \fi}

\begin{document}
%
\title{Support Recovery of Sparse Signals}
%
%
%

\author{Yuzhe~Jin,
        ~Young-Han~Kim,~\IEEEmembership{Member,~IEEE,}
        and~Bhaskar~D.~Rao,~\IEEEmembership{Fellow,~IEEE}\thanks{The material in this paper was presented in part at the IEEE International Conference on Acoustics, Speech, and Signal Processing (ICASSP), Las Vegas, Nevada, USA, March 2008 and the IEEE International Symposium on Information Theory (ISIT), Toronto, Ontario, Canada, July 2008. A short version of the paper was submitted to the IEEE International Symposium on Information Theory (ISIT), Austin, Texas, USA, 2010.}}
\maketitle

\begin{abstract}
We consider the problem of exact support recovery of sparse signals via noisy measurements. The main focus is the sufficient and necessary conditions on the number of measurements for support recovery to be reliable. By drawing
an analogy between the problem of support recovery and the problem of channel coding over the Gaussian multiple access channel, and exploiting mathematical tools developed for the latter problem, we obtain an information theoretic framework for analyzing the performance limits of support recovery. Sharp sufficient and necessary conditions on the number of measurements in terms of the signal sparsity level and the measurement noise level are derived. Specifically, when the number of nonzero entries is held fixed, the exact asymptotics on the number of measurements for support recovery is developed. When the number of nonzero entries increases in certain manners, we obtain sufficient conditions tighter than existing results. In addition, we show that the proposed methodology can deal with a variety of models of sparse signal recovery, hence demonstrating its potential as an effective analytical tool.
\end{abstract} 
\section{Introduction}
Consider the estimation of a sparse signal
$\Xv\in\mathbb{R}^m$ in high dimension via linear measurements
$\Yv = A\Xv+\Zv$, where $A\in\mathbb{R}^{n\times
  m}$ is referred to as the measurement matrix and $\Zv$ is the
measurement noise. A sparse signal is informally described as a signal whose representation in certain coordinates contains a large proportion of zero coefficients. In this paper, we mainly consider signals that are sparse with respect to the canonical basis of the Euclidean space. The goal is to estimate the sparse signal $\Xv$
by making as few number of measurements as possible. This problem has received much attention from many
research principles, motivated by a wide spectrum of applications such
as compressed sensing \cite{Donoho06_4,Candes_comp}, biomagnetic inverse problems \cite{Rao97}, \cite{Rao98}, image
processing \cite{Jeffs98}, \cite{Baraniuksinglepixel}, bandlimited
extrapolation and spectral estimation \cite{Parks91}, robust regression and outlier detection \cite{JR_10ICASSP}, speech
processing \cite{ChuSpeech}, channel estimation \cite{RaoCotter02},
\cite{NowakChannel}, echo cancellation \cite{Duttweiler2000}, \cite{RaoSong}, and
wireless communication \cite{RaoCotter02}, \cite{Guo09ITA}.

Computationally efficient algorithms for sparse signal recovery have been
proposed to find or approximate the sparse solution $\Xv$ in
various settings. A partial list includes matching pursuit \cite{Zhang93}, orthogonal
matching pursuit \cite{Pati93}, lasso \cite{Tibshirani94}, basis
pursuit \cite{Donoho01}, FOCUSS \cite{Rao97}, sparse Bayesian learning
\cite{Tipping01}, finite rate of innovation \cite{Vetterli02},
CoSaMP\cite{NeedellTropp08}, and subspace pursuit
\cite{DaiMilenkovic08}. At the same time, many exciting mathematical
tools have been developed to analyze the performance of these
algorithms. In particular, Donoho \cite{Donoho06_4}, Donoho, Elad, and Temlyakov \cite{Donoho06_1},
and Cand\`{e}s and Tao \cite{Tao05a}, and Cand\`{e}s, Romberg, and Tao\cite{Tao06} presented
sufficient conditions for $\ell_1$-norm minimization algorithms,
including basis pursuit, to successfully recover the sparse signals with respect to certain performance metrics. Tropp \cite{Tropp04}, Tropp and Gilbert \cite{Tropp07}, and Donoho, Tsaig, Drori, and Starck \cite{Donoho06} studied
greedy sequential selection methods such as matching pursuit
and its variants. In these papers, the structural properties of the
measurement matrix $A$, including coherence metrics
\cite{Zhang93}, \cite{Donoho06_1}, \cite{Tropp04}, \cite{Huo01} and
 spectral properties \cite{Donoho06_4}, \cite{Tao05a}, are
used as the major ingredient of the performance analysis. By using
random sensing matrices, these results translate to relatively simple
tradeoffs between the dimension of the signal $\Xv$, the number of nonzero entries in $\Xv$, and the number of measurements to ensure asymptotically successful reconstruction of the sparse
signal. In the absence of measurement noise, i.e., $\Zv = 0$,
the performance metric employed is the ability to recover the exact sparse signal \cite{Tao05a}. When the
measurement noise is present, the Euclidean distance between the
recovered signal and the true signal has been often employed as the performance metric \cite{Donoho06_1}, \cite{Tao06}.

In many applications, however, finding the exact support of the signal
is important even in the noisy setting. For example, in applications of medical imaging, magnetoencephalography (MEG) and electroencephalography (EEG) are common approaches for collecting noninvasive measurements of external electromagnetic signals \cite{Leahy_MEGEEG}. A relatively fine spatial resolution is required to localize the neural electrical activities from a huge number of potential locations \cite{WipfNaga2008}. In the domain of cognitive radio, spectrum sensing plays an important role in identifying available spectrum for communication, where estimating the number of active subbands and their locations becomes a nontrivial task \cite{Giannakis2007ICASSP}. In multiple-user communication systems such as a code-division multiple access (CDMA) system, the problem of neighbor discovery requires identification of active nodes from all potential nodes in a network based on a linear superposition of the signature waveforms of the active nodes \cite{Guo09ITA}. In all these problems, finding the support of the sparse signal is more important than approximating the signal vector in the Euclidean distance. Hence, it is important to understand performance issues in the exact support recovery of sparse signals with noisy measurements. Information theoretic tools have proven successful in this direction. Wainwright \cite{Wain07}, \cite{Wain09b} considered the problem of
exact support recovery using the optimal maximum likelihood decoder. Necessary
and sufficient conditions are established for different scalings
between the sparsity level and signal dimension. Using the same decoder, Fletcher, Rangan, and Goyal
\cite{Fletcher08NIPS}, \cite{Fletcher09IT} recently improved the necessary
condition. Wang, Wainwright, and Ramchandran \cite{Wang08ISIT} also presented a set of necessary conditions for exact support recovery. Ak\c{c}akaya and Tarokh \cite{Akcakaya} analyzed the
performance of a joint typicality decoder and applied it to find a set
of necessary and sufficient conditions under different performance
metrics including the one for exact support recovery. In addition, a
series of papers have leveraged many information theoretic
tools, including rate-distortion theory \cite{Bara06},
\cite{Fletcher_07}, expander graphs \cite{Hassibi09}, belief
propagation and list decoding \cite{Milenkovic09}, and low-density
parity-check codes \cite{Pfister08}, to design novel algorithms for
sparse signal recovery and to analyze their performances.

In this paper, we develop sharper asymptotic tradeoffs between the signal dimension $m$, the number of nonzero entries $k$, and number of measurements $n$ for reliable support recovery in the noisy setting. When $k$ is held fixed, we show that $ n = (\log m)/{c(\Xv)}$ is sufficient and necessary. We give a complete characterization of $c(\Xv)$ that depends on the values of the nonzero entries of $\Xv$. When $k$ increases in certain manners as specified later, we obtain sufficient and necessary conditions for perfect support recovery which improve upon existing results. Our main results are
inspired by the analogy to communication over the additive white
Gaussian noise multiple access channel (AWGN-MAC)
\cite{Jin08,Jin_ISIT}. According to this connection, the columns of
the measurement matrix form a common codebook for all
senders. Codewords from the senders are individually multiplied by
unknown channel gains, which correspond to nonzero entries of
$\Xv$. Then, the noise corrupted linear combination of these codewords
is observed. Thus, support recovery can be interpreted as decoding
messages from multiple senders. With appropriate modifications, the
techniques for deriving multiple-user channel capacity can be leveraged to provide
performance tradeoffs for support recovery.

The analogy between the problems of sparse signal recovery and channel coding has been observed from various perspectives in parallel work\cite{Bara06}, \cite[IV-D]{Tropp06}, \cite[II-A]{Wang08ISIT}, \cite[III-A]{Akcakaya}, \cite[11.2]{Donoho06}. However, our approach is different from the existing literature in several aspects. First, we explicitly connect the problem of exact support recovery to that of multiple access communication by interpreting
the sparse signal measurement model as a multiple access channel model. In spite of their similarity, however, there are also important differences between them which make a straightforward translation of known results nontrivial. We customize tools from multiple-user information theory (e.g., signal value estimation, distance decoding, Fano's inequality) to tackle the support recovery problem. Second, equipped with this analytic framework, we can obtain a performance tradeoff sharper than existing results. Moreover, the analytical framework can be extended to different models of sparse signal recovery, such as non-Gaussian measurement noise, sources with random activity levels, and multiple measurement vectors (MMV).

The rest of the paper is organized as follows.  We formally state the support recovery problem in Section \ref{sec2}. To motivate the main results of the paper and their proof techniques, we discuss in Section \ref{sec3} the similarities and differences between the support recovery problem and the multiple
access communication problem. Our main results are presented in Section \ref{sec:growingK}, together with comparisons to existing results in the literature.  The proofs of the main theorems are presented in Appendices \ref{sec4}, \ref{sec:Appen_Proof_Th2}, \ref{sec:ProofTh3}, and \ref{sec:Proof_Th4}, respectively. Section \ref{sec:moreextensions} further extends the results to different signal models and measurement procedures.

Throughout this paper, a set is a collection of unique objects. Let $\mathbb{R}^m$ denote the $m$-dimensional real
Euclidean space. Let $\mathbb{N}=\{1, 2, 3, ...\}$ denote the set of
natural numbers. Let $[k]$ denote the set $\{1, 2, ...,
k\}$. The notation $|S|$ denotes the cardinality of set $S$,
$\|\xv\|$ denotes the $\ell_2$-norm of a vector $\xv$,
and $\|A\|_F$ denotes the Frobenius norm of a matrix $A$.
The expression $f(x) = o(g(x))$ denotes $\lim_{x\rightarrow\infty}\frac{f(x)}{g(x)}=0$, $f(x) = O(g(x))$ denotes $|f(x)|\leq \alpha|g(x)|$ as $x\rightarrow\infty$ for some constant $\alpha>0$, $f(x) = \Theta(g(x))$ denotes $f(x) = O(g(x))$ and $g(x) = O(f(x))$, $f(x) = \Omega(g(x))$ denotes $g(x) = O(f(x))$, and $f(x) = \omega(g(x))$ denotes $g(x) = o(f(x))$.

\section{Problem Formulation}
\label{sec2}

Let $\wv=[w_1, ..., w_k]^\intercal\in\mathbb{R}^k$, where $w_i \neq 0$ for all $i$. Let $\Sv=[S_1, ..., S_k]^\intercal\in[m]^k$ be such that $S_1$, ..., $S_k$  are chosen uniformly at random from $[m]$ without replacement. In particular, $\{S_1, ...,S_k\}$ is uniformly distributed over all size-$k$ subsets of $[m]$. Then, the signal of interest $\Xv=\Xv(\wv,\Sv)$ is generated as
\begin{align}
X_s = \left\{ \begin{array}{ll}
w_j & \mbox{if $s=S_j$},\\
0 & \mbox{if $s\notin \{S_1, ..., S_k\}$}.\end{array} \right.
\label{signal_model}
\end{align}
Thus, the support of $\Xv$ is $\textmd{supp}(\Xv)=\{S_1, ..., S_k\}$. According to the signal model (\ref{signal_model}), $|\textmd{supp}(\Xv)|= k$. Throughout this paper, we assume $k$ is known. The signal is said to be sparse when $k\ll m$.

We measure $\Xv$ through the linear operation
\begin{align}
\Yv = A \Xv + \Zv
\label{CS_model}
\end{align}
where $A\in \mathbb{R}^{n\times m}$ is the measurement matrix, $\Zv\in\mathbb{R}^n$ is the measurement noise, and $\Yv\in\mathbb{R}^{n}$ is the noisy measurement. We further assume that the noise $Z_i$ are independently and identically distributed (i.i.d.) according to the Gaussian distribution $\mathcal{N}(0, \sigma_z^2)$.

Upon observing the noisy measurement $\Yv$, the goal is to recover the support of the sparse signal $\Xv$. A support recovery map is defined as
\begin{align}
d: \mathbb{R}^n \longmapsto 2^{[m]}.
\label{recon_algorithm}
\end{align}

Given the signal model (\ref{signal_model}), the measurement model (\ref{CS_model}), and the support recovery map (\ref{recon_algorithm}), we define the average probability of error by
\begin{align}
\overline{P}_e(\wv, A) \triangleq \P\{d(\Yv)\neq \textmd{supp}(\Xv(\wv, \Sv))\}
\label{average_prob}
\end{align}
for each (unknown) signal value vector $\wv\in\mathbb{R}^k$.


\section{An Information Theoretic Perspective on Sparse Signal Recovery}
\label{sec3}

In this section, we will introduce an important interpretation of the problem of sparse signal recovery via a communication problem over the Gaussian multiple access channel. The similarities and differences between the two problems will be elucidated, hence progressively unraveling the intuition and facilitating technical preparation for the main results and their proof techniques.

\subsection{Brief Review on the AWGN-MAC}
\label{sec:AWGN_MAC}
We start by reviewing the background on the $k$-sender multiple access channel (MAC). Suppose the senders wish to transmit information to a common receiver. Each sender $i$ has access to a codebook $\mathscr{C}^{(i)}=\{\mathbf{c}_1^{(i)}, \mathbf{c}_2^{(i)},..., \mathbf{c}_{m^{(i)}}^{(i)}\}$, where $\mathbf{c}_j^{(i)}\in\mathbb{R}^n$ is a codeword and $m^{(i)}$ is the number of codewords in $\mathscr{C}^{(i)}$. The rate for the $i$th sender is
$
R^{(i)} = (\log m^{(i)})/n
$.
To transmit information, each sender chooses a codeword from its codebook, and all senders transmit their codewords simultaneously over an AWGN-MAC \cite{Cover06}:
\begin{align}
Y_l =  h_1X_{1,l} + h_2X_{2,l}  + \cdots +h_k X_{k,l} + Z_l,\quad l=1, 2,...,n
\label{MAC_channel}
\end{align}
where $X_{i,l}$ denotes the input symbol from the $i$th sender to the channel at the $l$th use of the channel, $h_i$ denotes the channel gain associated with the $i$th sender, $Z_l$ is the additive noise, i.i.d. $\mathcal{N}(0, \sigma_z^2)$, and $Y_l$ is the channel output.

Upon receiving $Y_1,...,Y_n$, the receiver needs to determine the codewords transmitted by each sender. Since the senders interfere with each other, there is an inherent tradeoff among their operating rates. The notion of capacity region is introduced to capture this tradeoff by characterizing all possible rate tuples $(R^{(1)}, R^{(2)}, ..., R^{(k)})$ at which reliable communication can be achieved with diminishing error probability of decoding. By assuming each sender obeys the power constraint $
\|\mathbf{c}_j^{(i)}\|^2/n\leq \sigma_c^2
$
for all $j\in[m^{(i)}]$ and all $i\in\mathbb{N}_k$, the capacity region of an AWGN-MAC with known channel gains \cite{Cover06} is
\begin{align}
\left\{(R^{(1)},...,R^{(k)}):\sum_{i\in \Tc} R^{(i)} \leq \frac{1}{2}\log \left(1+\frac{\sigma_c^2}{\sigma_z^2}\sum\limits_{i\in \Tc} h_i^2
\right), \forall ~\Tc \subseteq [k]\right\}.
\label{Kusecaparegion}
\end{align}

\subsection{Connecting Sparse Signal Recovery to the AWGN-MAC}
\label{sec:connection_CS_MAC}
In the measurement model (\ref{CS_model}), one can remove the columns in $A$ which are nulled out by zero entries in $\Xv$ and obtain the following effective form of the measurement procedure
\begin{align}
\Yv =  X_{S_1} \mathbf{a}_{S_1} + \cdots + X_{S_k} \mathbf{a}_{S_k} + \Zv.
\label{CS_peeloff}
\end{align}
By contrasting (\ref{CS_peeloff}) to AWGN MAC (\ref{MAC_channel}), we can draw the following key connections that relate the two problems \cite{Jin08}.

\begin{enumerate}
\item {\bf A nonzero entry as a sender}: We can view the existence of a nonzero entry position $S_j$ as sender $j$ that accesses the MAC.

\item  {\bf  $\mathbf{a}_j$ as a codeword}: We treat the measurement matrix $A$ as a codebook with each column
$\mathbf{a}_j$, $j\in [m]$, as a codeword. Each element of $\mathbf{a}_{S_i}$ is fed one by one to the channel (\ref{MAC_channel}) as the input symbol $X_i$, resulting in $n$ uses of the channel. The noise $\Zv$ and  measurement $\Yv$ can be related to the channel noise $Z$ and channel output $Y$ in the same fashion.

\item {\bf $X_{S_i}$ as a channel gain}: The nonzero entry $X_{S_i}$ in (\ref{CS_peeloff}) plays the role of the channel gain $h_i$ in (\ref{MAC_channel}). Essentially, we can interpret the vector representation (\ref{CS_peeloff}) as $n$ consecutive uses of the $k$-sender AWGN-MAC (\ref{MAC_channel}) with appropriate stacking of the inputs/outputs into vectors.

\item {\bf Similarity between objectives}: In the problem of sparse signal recovery, the goal is to find the support $\{S_1,...,S_k\}$ of the signal. In the problem of MAC communication, the receiver's goal is to determine the indices of codewords, i.e., $S_1,...,S_k$, that are transmitted by the senders.
\end{enumerate}

Based on the abovementioned aspects, the two problems share significant similarities which enable leveraging the information theoretic methods for performance analysis of support recovery of sparse signals. However, as we shall see next, there are domain specific differences between the support recovery problem and the channel coding problem that should be addressed accordingly to rigorously apply the information theoretic approaches.

\subsection{Key Differences}
\label{Section:difference}
\begin{enumerate}
\item\textbf{Common codebook}: In MAC communication, each sender uses its own codebook. However, in sparse signal recovery, the ``codebook'' $A$ is shared by all ``senders''. All senders choose their codewords from the same codebook and hence operate at the same rate. Different senders will not choose the same codeword, or they will collapse into one sender.

\item\textbf{Unknown channel gains}: In MAC communication, the capacity region (\ref{Kusecaparegion}) is valid assuming that the receiver knows the channel gain $h_i$ \cite{Tse05}. In contrast, for sparse signal recovery problem, $X_{S_i}$ is actually unknown and needs to be estimated. Although coding techniques and capacity results are available for communication with channel uncertainty, a closer examination indicates that those results are not directly applicable to our problem. For instance, channel training with pilot symbols is a common practice to combat channel uncertainty \cite{Hassibi00howmuch}. However, it is not obvious how to incorporate the training procedure into the measurement model (\ref{CS_model}), and hence the related results are not directly applicable.
\end{enumerate}

Once these differences are properly accounted for, the connection between the problems of sparse signal recovery and channel coding makes available a variety of information theoretic tools for handling performance issues pertaining to the support recovery problem. Based on techniques that are rooted in channel capacity results, but suitably modified to deal with the differences, we will present the main results of this paper in the next section.

\section{Main Results and Their Implications}
\label{sec:growingK}

\subsection{Fixed Number of Nonzero Entries}
To discover the precise impact of the values of the nonzero entries on support recovery, we consider the support recovery of a sequence of sparse signals generated with the same signal value vector $\wv$. In particular, we assume that $k$ is fixed. Define the auxiliary quantity
\begin{align}
c(\wv) \triangleq \min_{\Tc \subseteq [k]} \left[\frac{1}{2|\Tc|} \log
\left(1+\frac{\sigma_a^2}{\sigma_z^2}\sum\limits_{j\in \Tc } w_j^2
\right)\right].
\label{def_C}
\end{align}
For example, when $k=2$,
\begin{align}
c(w_1, w_2) = \min \left[ \frac{1}{2} \log
\left(1+\frac{\sigma_a^2 w_1^2}{\sigma_z^2} \right), \frac{1}{2} \log
\left(1+\frac{\sigma_a^2 w_2^2}{\sigma_z^2} \right), \frac{1}{4} \log
\left(1+\frac{\sigma_a^2 (w_1^2+w_2^2)}{\sigma_z^2} \right)\right].\nonumber
\end{align}
We can see from Section \ref{sec3} that this quantity is closely related to the $2$-sender multiple access channel capacity with equal-rate constraint.

In the following two theorems, we summarize our main results under the assumption that $k$ is fixed. The subscript in $n_m$ denotes possible dependence between $n$ and $m$. The proof of the theorems are presented in Appendices \ref{sec4} and \ref{sec:Appen_Proof_Th2}, respectively.

\begin{theorem}
If
\begin{align}
\limsup_{m\rightarrow \infty} \frac{\log m}{n_m} < c(\wv)
\label{condition_theorem1}
\end{align}
then there exist a sequence of matrices $\{A^{(m)}\}_{m=k}^\infty$, $A^{(m)}\in\mathbb{R}^{n_m\times m}$, and a sequence of support recovery maps $\{d^{(m)}\}_{m=k}^\infty$, $d^{(m)}:\mathbb{R}^{n_m} \mapsto 2^{[m]}$, such that
\begin{align}
\frac{1}{n_m m}\|A^{(m)}\|_F^2 \leq \sigma_a^2
\label{Power_cons}
\end{align}
and
\begin{align}
\lim_{m\rightarrow\infty} \overline{P}_e(\wv,A^{(m)}) = 0.
\end{align}
\end{theorem}

\begin{theorem}
If there exist a sequence of matrices $\{A^{(m)}\}_{m=k}^\infty$, $A^{(m)}\in\mathbb{R}^{n_m\times m}$, and a sequence of support recovery maps $\{d^{(m)}\}_{m=k}^\infty$, $d^{(m)}:\mathbb{R}^{n_m}\mapsto 2^{[m]}$, such that
\begin{align}
\frac{1}{n_m m}\|A^{(m)}\|_F^2 \leq \sigma_a^2
\end{align}
and
\begin{align}
\lim_{m\rightarrow\infty} \overline{P}_e(\wv,A^{(m)}) = 0
\end{align}
then
\begin{align}
\limsup_{m\rightarrow \infty} \frac{\log m}{n_m} \leq c(\wv).
\end{align}
\end{theorem}

\medskip

Theorems 1 and 2 together indicate that $n = (\log m) / (c(\wv) \pm \epsilon)$ is sufficient and necessary for exact support recovery. The constant $c(\wv)$ is explicitly characterized, capturing the role of signal strength in support recovery.


\subsection{Growing Number of Nonzero Entries}
Next, we consider the support recovery for the case where $k$, the number of nonzero entries, grows with $m$, the dimension of the signal. We assume that the magnitude of a nonzero entry is bounded from both below and above. Meanwhile, we consider using random measurement matrices drawn from the Gaussian distribution, which makes it more convenient to compare with existing results in the literature. Note that we can easily establish corresponding results on the existence of arbitrary measurement matrices as in Theorems 1 and 2.

First, we present a sufficient condition for exact support recovery. The proof can be found in Appendix \ref{sec:ProofTh3}.
\begin{theorem}
Let $\{\wv^{(m)}\}_{m=1}^\infty$ be a sequence of vectors satisfying
$\wv^{(m)}\in\mathbb{R}^{k_m}$ and $0<w_\text{min}\leq|w_j^{(m)}| \leq w_{\text{max}}<\infty$ for all $j\in[k_m], m\geq 1$. Let $A^{(m)}\in\mathbb{R}^{n_m\times m}$ be generated as $A_{ij}^{(m)}\sim\mathcal{N}(0, \sigma_a^2)$. If
\begin{align}
\limsup_{m\rightarrow \infty}  \frac{1}{n_m} \max_{j\in[k_m]}\left[\frac{6k_m\log k_m + 2j\log m}{\log\left(\frac{j w_{\text{min}}^2\sigma_a^2}{\sigma_z^2}+1\right)}\right] < 1
\label{growingK_suff}
\end{align}
then there exists a sequence of support recovery maps $\{d^{(m)}\}_{m=1}^\infty$, $d^{(m)}:\mathbb{R}^{n_m} \mapsto 2^{[m]}$, such that
\[
\lim_{m\rightarrow \infty} \P\{d^{(m)}(A\Xv(\wv^{(m)}, \Sv) + \Zv)\neq \textmd{supp}(\Xv(\wv^{(m)}, \Sv))\}=0.\nonumber
\]
\end{theorem}
\medskip

To better understand Theorem 3, we present the following implication of (\ref{growingK_suff}) that shows the tradeoffs between the order of $n$ versus $m$ and $k$.
\begin{corollary}
Under the assumption of Theorem 3,
\[
\lim_{m\rightarrow \infty} \P\{d^{(m)}(A\Xv(\wv^{(m)}, \Sv) + \Zv)\neq \textmd{supp}(\Xv(\wv^{(m)}, \Sv))\}=0\nonumber
\]
provided that
\[n = \max\left\{\Omega(k\log k),\Omega \left({\frac{k}{\log k}\log m}\right)\right\}.\]
In particular, we have the following:
\begin{enumerate}
\item When $m=k^{\Omega(\log k)}$, for example $m=k^{\log k}$, the sufficient number of of measurements is $n=\Omega(\frac{k\log m}{\log k})$.

\item When $e^{\omega(\log k)} \leq m \leq k^{o(\log k)}$, for example $m=k^{\log \log k}$, the sufficient number of of measurements is $n=\Omega(k\log k)$. In this case, $\log m = \omega (\log k)$, and hence $\Omega(k\log k) \leq \Omega(k\log m)$. Thus, $n=\Omega(k\log k)$ is a better sufficient condition than $n=\Omega(k\log m)$.

\item When $\omega(k) \leq m \leq e^{\Theta(\log k)}$, for example $m=k^2$, the sufficient number of of measurements is $n=\Omega(k\log m)$.

\item When $m=\Theta (k)$, the sufficient number of of measurements is $n=\Omega(k\log m)$.
\end{enumerate}

\end{corollary}
\medskip

The following table on the next page summarizes the sufficient orders of $n$ paired with different relations between $m$ and $k$ in Corollary 1.

\newpage

\begin{table}[t]
\renewcommand{\arraystretch}{1.5}
\centering
\begin{tabular}{c|c|c}
\hline
\multicolumn{2}{c|} {Relation between $m$ and $k$} & Sufficient $n$\\
\hline
\hline
\multirow{3}{*}{$m=\omega(k)$} &$m=k^{\Omega(\log k)}$  & $n=\Omega(\frac{k\log m}{\log k})$\\\cline{2-3}
& $e^{\omega(\log k)} \leq m \leq k^{o(\log k)}$  & $n=\Omega(k\log k)$\\\cline{2-3}
& $\omega(k) \leq m \leq e^{\Theta(\log k)}$ & $n=\Omega(k\log m)$\\
\hline
\hline
\multicolumn{2}{c|} {$m=\Theta (k)$} & $n=\Omega(k\log m)$\\
\hline
\end{tabular}
\end{table}

In the existing literature, Wainwright \cite{Wain09b} and Ak\c{c}akaya
and Tarokh \cite{Akcakaya} both derived sufficient conditions for exact
support recovery. Under the same assumption of Theorem 3, the
sufficient conditions presented in these papers, respectively, are
summarized in the following table:
\begin{table}[h]
\renewcommand{\arraystretch}{1.3}
\centering
\begin{tabular}{c|c|c}
\hline
Relation between $m$ and $k$ & Wainwright\cite{Wain09b} & Ak\c{c}akaya et al. \cite{Akcakaya}\\
\hline
\hline
$m=\omega(k)$ & $n = \Omega(k\log \frac{m}{k})$ & $n = \Omega(k\log (m-k))$\\
\hline
{$m=\Theta (k)$} & $n = \Omega(m)$ & $ n = \Omega(m)$\\
\hline
\end{tabular}
\end{table}

To compare the results, we first examine the case of $m=\omega(k)$
(i.e., sublinear sparsity). Note that in the regime where
$m=e^{\omega(\log k)}$, our sufficient condition on $n$ includes lower
order growth rate, hence is better, than existing results. In the
regime where $\omega(k)\leq m \leq e^{\Theta(\log k)}$, there exists a
certain scenario, e.g., $k=\frac{m}{\log m}$, in which our sufficient
condition is of the same order as in \cite{Akcakaya} but higher than
in \cite{Wain09b}. In the case of $m=\Theta(k)$ (i.e., linear
sparsity), we see that our sufficient condition is stricter, implying
its inferiority to existing results in this regime.

Next, we present a necessary condition, the proof of which can be found in Appendix \ref{sec:Proof_Th4}.
\begin{theorem}
Let $\{\wv^{(m)}\}_{m=1}^\infty$ be a sequence of vectors satisfying
$\wv^{(m)}\in\mathbb{R}^{k_m}$ and $0<w_\text{min}\leq|w_j^{(m)}| \leq w_{\text{max}}<\infty$ for all $j\in[k_m], m\geq 1$. Let $A^{(m)}\in\mathbb{R}^{n_m\times m}$ be generated as $A_{ij}^{(m)}\sim\mathcal{N}(0, \sigma_a^2)$. If
\begin{align}
\limsup_{m\rightarrow\infty}\frac{2k_m \log ({m}/{k_m}) }{n_m\log\left( \frac{2k_m w_{\textrm{max}}^2\sigma_a^2}{\sigma_z^2} +1\right)} > 1
\end{align}
then for any sequence of support recovery maps $\{d^{(m)}\}_{m=1}^\infty$, $d^{(m)}:\mathbb{R}^{n_m} \mapsto 2^{[m]}$, we have
\[
\liminf_{m\rightarrow \infty} \P\{d^{(m)}(A\Xv(\wv^{(m)}, \Sv) + \Zv)\neq \textmd{supp}(\Xv(\wv^{(m)}, \Sv))\} > 0.\nonumber
\]
\end{theorem}
\medskip

To compare with existing results under the same assumption\footnote{The necessary conditions derived in \cite{Wain09b}, \cite{Wang08ISIT}, and \cite{Akcakaya} were originally derived under slightly different assumptions. Here we adapted them to compare the asymptotic orders of $n$.} of Theorem 4, we first note that when $m=\Theta(k)$, the necessary condition is $n=\Omega(m)$, which follows simply from the elementary constraint $n \ge k$ that the number of measurements has to be no smaller than the number of nonzero entries for support recovery to be
possible. Contrasted by the sufficient conditions derived in \cite{Wain09b} and \cite{Akcakaya}, $n=\Omega(m)$ is the necessary and sufficient condition for linear sparsity.  When $m=\omega(k)$, we summarize the necessary conditions developed in previous papers in the following table:

\addtocounter{footnote}{1}
\footnotetext[\value{footnote}]{This result is implied in \cite{Akcakaya}, by identifying $C'_4$ in Thm. 1.6 therein, and clarifying the order of $n$. The proof of Thm. 1.6 states that (below its (25)) asymptotically reliable support recovery is not possible if $n< [\log(1+{\|\wv^{(m)}\|^2}/{\sigma_z^2})]^{-1}mH(k/m)-\log(m+1)$. Note that $mH(k/m) = \Theta(k\log(m/k))$. Hence, we consider $n= \Omega(\frac{k\log(m/k)}{\log k})$ an appropriate necessary condition resulting from the proof in \cite{Akcakaya}.}

\begin{table}[h]
\renewcommand{\arraystretch}{1.5}
\centering
\begin{tabular}{c||c}
\hline
Relation between $m$ and $k$ & $m=\omega(k)$\\
\hline
\hline
Wainwright \cite{Wain09b}  & $n = \Omega(\log m)$ \\
\hline
Wang et al. \cite{Wang08ISIT} & $n = \Omega(\frac{k\log (m/k)}{\log k})$ \\
\hline
Ak\c{c}akaya et al. \cite{Akcakaya}$^{\textrm{\decimal{footnote}}}$ & $n = \Omega(\frac{k\log(m/k)}{\log k})$ \\
\hline
Theorem 4 & $n =\Omega(\frac{k\log (m/k)}{\log k})$ \\
\hline
\end{tabular}
\end{table}


In this case, $n =\Omega(\frac{k\log (m/k)}{\log k})$ is the best known necessary condition.

\subsection{Further Observations}
Note that for the sublinear sparsity with $m=k^{\Omega(\log k)}$, both $\log\frac{m}{k}$ and $\log m$ are of the same order and hence our sufficient and necessary conditions both indicate $n =\Omega(\frac{k}{\log k}\log m)$. This provides a sharp performance tradeoff for support recovery in this specific regime, which to our knowledge has not been observed in previous work (see, for example, the remarks in \cite[III-A]{Wain09b}, \cite[III-Remark 2)]{Fletcher09IT}). For the regime where $\omega(k)\leq m\leq k^{o(\log k)}$, the orders of $n$ in any pair of sufficient and necessary conditions have a nontrivial difference, leaving an open question on further narrowing the gap in this remaining regime of sublinear sparsity.

In addition, it is worthwhile to note that our analytical framework could also be adapted to the case where $w_{\textrm{min}} = O(1/\sqrt{k})$. This is a scenario extensively discussed in \cite{Wain09b,Fletcher09IT,Wang08ISIT}. We will not pursue this direction in detail.

\section{Extensions}
\label{sec:moreextensions}
The connection between the problems of support recovery and channel coding can be further explored to provide the performance tradeoff for different models of sparse signal recovery. Next, we discuss its potential to address several important variants.

\subsection{Non-Gaussian Noise}
Note that the rules for support recovery, mainly reflected in (\ref{rec_rule}) and (\ref{C3_testing}) in the proof of Theorem 1 in Appendix \ref{sec4}, are similar to the method of nearest neighbor decoding in information theory. Following the argument in \cite{Lapidoth96}, one can show that by replacing the assumption in (\ref{CS_model}) on measurement noise $Z_i\sim\mathcal{N}(0, \sigma_z^2)$ with $\text{Var}(Z_i) = \sigma_z^2$, the results in the previous theorems continue to hold.

\subsection{Random Signal Activities}
In Theorem 1, $\wv$ is assumed to be a fixed vector of nonzero entries. We now relax this condition to allow random $\Wv$, which leads to sparse signals whose nonzero entries are randomly generated and located. For simplicity of exposition, assume that $k$ is fixed. Interestingly, the model (\ref{CS_model}) with this new assumption can now be contrasted to a MAC with random channel gains
\begin{align}
Y_l =  H_1 X_{1,l} + H_2 X_{2,l}  + \cdots + H_k X_{k,l}+Z_l, ~l=1, 2, ..., n.\label{MAC_random_channel}
\end{align}
The difference between (\ref{MAC_random_channel}) and (\ref{MAC_channel}) is that the channel gains $H_i$ are random variables in this case. Specifically, in order to contrast the problem of support recovery of sparse signals, $H_i$ should be considered as being realized once and then kept fixed during the entire channel use \cite{Jin08}. This channel model is usually termed as a slow fading channel \cite{Tse05}.

The following theorem states the performance of support recovery of sparse signals under random signal activities.

\begin{theorem} Suppose $\Wv$ has bounded support, and $\limsup_{m\rightarrow\infty}\frac{\log m}{n_m} = r.$ Let $A^{(m)}\in\mathbb{R}^{n_m\times m}$ be generated as $A_{ij}^{(m)}\sim\mathcal{N}(0, \sigma_a^2)$. Then, there exists a sequence of support recovery maps $\{d^{(m)}\}_{m=k}^\infty$, $d^{(m)}:\mathbb{R}^{n_m} \mapsto 2^{[m]}$, such that
\[
\limsup_{m\rightarrow\infty} {\P}\{d^{(m)}(A^{(m)}\Xv(\Wv,\Sv)+\Zv) \neq \textrm{supp}(\Xv)\} \leq \P\{c(\Wv) \leq r\}
\]
where $c(\Wv)$ is defined in (\ref{def_C}).
\end{theorem}
\medskip

\begin{IEEEproof} Note that
\begin{align}
&  \limsup_{m\rightarrow\infty} {\P}\{d^{(m)}(A^{(m)}\Xv(\Wv,\Sv)+\Zv) \neq \textrm{supp}(\Xv)\} \nonumber\\
&= \limsup_{m\rightarrow\infty} \int_{\wv}{\P}\{d^{(m)}(A^{(m)}\Xv(\wv,\Sv)+\Zv) \neq \textrm{supp}(\Xv)\} dF(\wv) \nonumber\\
&= \limsup_{m\rightarrow\infty}\int_{\wv: c(\wv)>r}{\P}\{d^{(m)}(A^{(m)}\Xv(\wv,\Sv)+\Zv) \neq \textrm{supp}(\Xv)\}dF(\wv) \nonumber\\
&~~~+ \limsup_{m\rightarrow\infty}\int_{\wv: c(\wv)\leq r}{\P}\{d^{(m)}(A^{(m)}\Xv(\wv,\Sv)+\Zv) \neq \textrm{supp}(\Xv)\}dF(\wv)\nonumber\\
&\leq  \int_{\wv: c(\wv)>r}\limsup_{m\rightarrow\infty}{\P}\{d^{(m)}(A^{(m)}\Xv(\wv,\Sv)+\Zv) \neq \textrm{supp}(\Xv)\}dF(\wv) + \int_{\wv: c(\wv)\leq r}dF(\wv) \label{fatou}\\
&\leq \P\{c(\Wv)\leq r\}\label{afterfatou}
\end{align}
where (\ref{fatou}) follows from Fatou's lemma \cite{Resnick} and (\ref{afterfatou}) follows by applying the proof of Theorem 1 to the integrand.
\end{IEEEproof}
\medskip

Theorem 5 implies that generally, rather than having a diminishing error probability, we have to tolerate certain error probability which is upper-bounded by $\P(c(\Wv)\leq r)$, when the nonzero values are randomly generated. Conversely, in order to design a system with probability of success at least $(1-p)$, one can find $r$ that satisfies $\P(c(\Wv)\leq r) \leq p$.
\medskip

\subsection{Multiple Measurement Vectors}
Recently, increasing research effort has been focused on sparse signal recovery with multiple measurement vectors (MMV) \cite{Rao05,Wipf_2007_b,JieChen,Cichocki,Eldar_UnionSubspace}. In this problem, we wish to measure multiple sparse signals $\Xv_1(\wv_1,\Sv)$, $\Xv_2(\wv_2,\Sv)$, $...$, and $\Xv_t(\wv_t,\Sv)$ that possess a common sparsity profile, that is, the locations of nonzero entries are the same in each $\Xv_t$. We use the same measurement matrix $A\in\mathbb{R}^{n\times m}$ to perform
\begin{align}
Y = AX + Z\label{Ext_MMV}
\end{align}
where $X = [\Xv_1, \Xv_2, ...,\Xv_t]\in\mathbb{R}^{m\times t}$, $Z=[\Zv_1, \Zv_2, ...,\Zv_t]\in\mathbb{R}^{n\times t}$ is the measurement noise, and $Y=[\Yv_1, \Yv_2, ...,\Yv_t]\in \mathbb{R}^{n\times t}$ is the noisy measurement.

Note that the model (\ref{CS_model}) can be viewed as a special case of the MMV model (\ref{Ext_MMV}) with $t=1$. The methodology that has been developed in this paper has the potential to be extended to deal with the performance issues with the MMV model by noting the following connections to channel coding\cite{Jin08}. First, the same set of columns in $A$ are scaled by entries in different $\Xv_j$, forming outputs as elements in different $\Yv_j$. The nonzero entries of $X$ can then be viewed as the coefficients that connect different pairs of inputs and outputs of a channel. Second, each measurement vector $\Yv_j$ can be viewed as the received symbols at the $j$th receiver, and hence the MMV model indeed corresponds to a multiple-input multiple-output (MIMO) channel model. Third, the aim is to recover the locations of nonzero rows of $X$ upon receiving $Y$. This implies that, in the language of MIMO channel communication, the receivers will fully collaborate to decode the information sent by all senders. Via proper accommodation of the method developed in this paper, the capacity results for MIMO channels can be leveraged to shed light on the performance tradeoff of sparse signal recovery with MMV.

\section{Concluding Remarks}
\label{sec6}
In this paper, we developed techniques rooted in multiple-user information theory to address the performance issues in the exact support recovery of sparse signals, and discovered necessary and sufficient conditions on the number of measurements. It is worthwhile to note that the interpretation of sparse signal recovery as MAC communication opens new avenues to different theoretic and algorithmic problems in sparse signal recovery. We conclude this paper by briefly discussing several interesting potential directions made possible by this interpretation:

\begin{enumerate}
\item Among the large collection of algorithms for sparse signal recovery, the sequential selection methods, including matching pursuit \cite{Zhang93} and orthogonal matching pursuit (OMP) \cite{Pati93}, determine one nonzero entry at a time, remove its contribution in the residual signal, and repeat this procedure until certain stopping criterion is satisfied. In contrast, the class of convex relaxation methods, including basis pursuit \cite{Donoho01} and lasso \cite{Tibshirani94}, jointly estimate the nonzero entries.

    Potentially, the sequential selection methods can be viewed as successive interference cancellation (SIC) decoding \cite{Tse05} for multiple access channels, whereas the convex relaxation methods can be viewed as joint decoding. It would be interesting to ask whether one can make these analogies more precise and use them to address performance issues. Similarities at an intuitive level between OMP and SIC have been discussed in \cite{Jin_ISIT} with performance results supported by empirical evidence. More insights are yet to be explored.

\item The design of channel codes and the development of decoding methods have been extensively studied in the contexts of information theory and wireless communication. Some of these ideas have been transformed into design principles for sparse signal recovery \cite{Hassibi09}, \cite{Milenkovic09}, \cite{Pfister08} as mentioned in the introduction. Thus far, however, the efforts in utilizing the codebook designs and decoding methods are mainly focused on the point-to-point channel model, which implies that the recovery methods iterate between first recovering one nonzero entry or a group of nonzero entries by treating the rest of them as noise and then removing the recovered nonzero entries from the residual signal. In this paper, we established the analogy between the sparse signal recovery and the multiple access communication. It motivates us to envision opportunities beyond a point-to-point channel model. As one important question, for example, can we develop practical codes for joint decoding and reconstruction techniques to simultaneously recover all the nonzero entries?

\item Last but not the least, we return to one remaining open question from this paper. Recall that for sublinear sparsity, there exists a certain regime in which the tight bound on the number of measurements is not known yet. Can we further improve the result in this regime, thereby closing the gap between sufficient and necessary conditions on the number of measurements for arbitrary scalings among the model parameters?
\end{enumerate}

\section{Acknowledgements}
This research was supported by NSF Grants CCF-0830612 and CCF-0747111. Y. J. wants to thank Liwen Yu for insightful discussions on perspectives of MAC communication.

\appendices
\section{Proof of Theorem 1}
\label{sec4}
The proof of Theorem 1 employs the distance decoding technique \cite{Lapidoth96}. We will first randomly generate the measurement matrix $A^{(m)}$ and show that the error probability averaged over this ensemble tends to zero as $m\rightarrow \infty$. This naturally leads to the existence of a sequence of deterministic matrices for achieving diminishing error probability of support reconstruction. We randomly generate the measurement matrix $A^{(m)}$ with entries
drawn independently according to $A_{ij}^{(m)}\sim \mathcal{N}(0,
\sigma_a^2)$, $i\in[n_m], j\in [m]$.  Let
$\Av_j^{(m)}$ denote the $j$th column of $A^{(m)}$.

For simplicity of exposition, we describe the support recovery procedure
for two distinct cases on the number of nonzero entries.

\textbf{Case 1}: $k=1$. In this case, the signal of interest is $\Xv =
\Xv(w_1, S_1)$. Consider the following support recovery procedures.
Fix $\e > 0$. First form an estimate $\hat{W}$ of $|w_1|$ as
\begin{align}
\hat{W} \triangleq \sqrt{\frac{\left|\frac{1}{n_m}\|\Yv\|^2 - \sigma_z^2\right|}{\sigma_a^2}}.
\label{C1_est_gain}
\end{align}
Declare that $\sh_1 \in [m]$ is the estimated
location for the nonzero entry, i.e., $d^{(m)}(\Yv) = \{\sh_1\}$, if it is the unique index such that
\begin{align}
\frac{1}{n_m}\|\Yv - (-1)^q \hat{W} {\Av}_{\sh_1}^{(m)} \|^2 \leq
\sigma_z^2 + \epsilon^2 \sigma^2_a
\label{rec_rule}
\end{align}
for either $q = 1$ or $q = 2$. If there is none or more than one, pick
an arbitrary index.

We now analyze the average probability of error
\[
\P(\Ec) = \E[\overline{P}_e(w_1, A^{(m)})]
= \P\{d^{(m)}(\Yv)\neq \{S_1\}\},
\]
where the expectation is taken with respect to the random measurement matrix $A^{(m)}$.  Due to the
symmetry in the problem and the measurement matrix generation, we assume without loss of generality $S_1 = 1$, that is,
\begin{align}
\Yv = w_1\Av_1^{(m)}+\Zv\nonumber
\end{align}
for some $w_1$.  In the following analysis, we drop superscripts and
subscripts on $m$ for notational simplicity when no ambiguity
arises. Define the events
\begin{align}
\Ec_{s} \triangleq \left\{\exists q\in\{1,2\} \text{ such that }
\frac{1}{n}\|\Yv - (-1)^q\hat{W} \Av_s \|^2 \leq \sigma_z^2
+ \epsilon^2  \sigma^2_a\right\}, ~s\in[m].\nonumber
\end{align}
Then
\begin{align}
\P(\Ec) \le \P\left(\Ec_{1}^c \cup\left(\cup_{s=2}^{m} \Ec_{s} \right)\right).
\label{error_events}
\end{align}
Let
\begin{align}
\Ec_\text{aux} &\triangleq \left\{\hat{W}-|w_1| \in (-\epsilon, \epsilon)\right\}  \cap \left\{\frac{1}{n}\|\Yv\|^2 - [w_1^2 \sigma_a^2 + \sigma_z^2] \in (-\epsilon, \epsilon)\right\}.\nonumber
\end{align}
Then, by the union of events bound and the fact that $\Ac^c\cup\Bc =
\Ac^c\cup (\Bc \cap \Ac)$,
\begin{align}
\P(\Ec)
&\leq \P(\Ec_\text{aux}^c) + \P(\Ec_1^c) + \sum_{s=2}^m \P(\Ec_s \cap \Ec_\text{aux}).
\label{C1_Error}
\end{align}

We bound each term in (\ref{C1_Error}). First, by the weak law of large numbers (LLN), $\lim_{m\rightarrow \infty}
\P(\Ec_\text{aux}^c)=0.$ Next, we consider $\P(\Ec_1^c)$. If $w_1>0$,
\begin{align}
\frac{1}{n}\|\Yv - \hat{W} \Av_1 \|^2 = \frac{1}{n}\| w_1\Av_1 + \Zv - \hat{W} \Av_1 \|^2 = (w_1-\hat{W})^2\frac{\|\Av_1\|^2}{n} + 2(w_1-\hat{W})\frac{\Av_1^\intercal \Zv}{n} + \frac{\|\Zv\|^2}{n}.
\label{E_12c}
\end{align}
For any $\epsilon_1>0$, as $m\rightarrow \infty$, by the LLN,
\[
\P\bigg(\left\{w_1-\hat{W} \in (-\epsilon_1, \epsilon_1)\right\} \cap \left\{\frac{\|\Av_1\|^2}{n} - \sigma_a^2 \in (-\epsilon_1, \epsilon_1)\right\}\bigg) \rightarrow 1.
\label{event_term1}
\]
Hence, we have for the first term in (\ref{E_12c})
\[
\P\left((w_1-\hat{W})^2 \frac{\|\Av_1\|^2}{n}\in [0, \epsilon_1^2 \sigma_a^2+\epsilon_1^3)\right) \rightarrow 1.
\]
Following a similar reasoning, for the second term in (\ref{E_12c}),
\[
\P\bigg((w_1-\hat{W}) \frac{\Av_1^\intercal\Zv}{n} \in (-\epsilon_1^2, \epsilon_1^2)\bigg) \rightarrow 1
\]
and for the third term,
\[
\P\bigg(\frac{\|\Zv\|^2}{n} \in (\sigma_z^2-\epsilon_1, \sigma_z^2+\epsilon_1)\bigg) \rightarrow 1.
\]
Therefore, for any $\epsilon_1 >0$,
\[
\lim_{m\rightarrow\infty}\P\left(\frac{1}{n}\|\Yv - \hat{W} \Av_1 \|^2 \in(\sigma_z^2-\epsilon_1, \sigma_z^2+\epsilon_1) \right)=1
\]
which implies that
\[
\lim_{m\rightarrow\infty}\P\left(\frac{1}{n}\|\Yv - \hat{W} \Av_1 \|^2 \leq \sigma_z^2+\epsilon^2\sigma_a^2 \right)=1.
\]
Similarly, if $w_1<0$,
\[
\lim_{m\rightarrow\infty}\P\left(\frac{1}{n}\|\Yv + \hat{W} \Av_1 \|^2 \leq \sigma_z^2+\epsilon^2\sigma_a^2 \right)=1.
\]
Hence, $\lim_{m\rightarrow\infty}\P(\Ec_1^c) = 0.$

For the third term in (\ref{C1_Error}), we need the following lemma, whose proof is presented at the end of this appendix:

\begin{lemma}
\label{lemma1}
Let $0<\beta<\alpha$. Let $\{u_i\}_{i=1}^{n}$ be a real sequence satisfying
\[
\frac{1}{n}\sum_{i=1}^n {u}_i^2 \in (\alpha-\beta, \alpha+\beta).\nonumber
\]
Let $\{V_i\}_{i=1}^{n}$ be an i.i.d. random sequence where $V_i \sim \mathcal{N}(0, \sigma_v^2)$. Then, for any $\gamma\in(0, \alpha-\beta)$,
\[
\P\left(\frac{1}{n}\sum_{i=1}^n(u_i-V_i)^2 \leq \gamma \right)
\leq 2^{-\frac{n}{2}\log\left(\frac{\alpha-\beta}{\gamma}\right)}.
\]
\end{lemma}

\medskip

Continuing the proof of Theorem 1, we consider $\P(\Ec_s\cap\Ec_\text{aux})$ for $s \ne 1$. Then
\begin{align*}
\P(\Ec_s\cap\Ec_\text{aux}) \leq \P(\Ec_s|\Ec_\text{aux}) =\int_{\yv\in \Ec_\text{aux}} \P(\Ec_s|\{\Yv =\yv\} \cap \Ec_\text{aux})f(\yv|\Ec_\text{aux})d\yv.\nonumber
\end{align*}
Since $\Av_s$ is independent of $\Yv$ and $\hat{W}$, it
follows from the definition of $\Ec_\text{aux}$ and Lemma~1 (with $\alpha = w_1^2 \sigma_a^2 + \sigma_z^2$ and $\gamma = \sigma_z^2 + \epsilon^2 \sigma^2_a$) that
\[
\P\left(\frac{1}{n}\|\Yv - (-1)^q \hat{W} \Av_s \|^2\leq \sigma_z^2 + \epsilon^2 \sigma^2_a
\;\Big|\; \{\Yv =\yv\} \cap\Ec_\text{aux}\right)\leq  2^{-\frac{n}{2}\log\left(\frac{w_1^2 \sigma_a^2 + \sigma_z^2-\epsilon}{\sigma_z^2 + \epsilon^2 \sigma^2_a} \right)}\nonumber
\]
for $q = 1,2$, if $\epsilon$ is sufficiently small. Thus,
\[
\P(\Ec_s|\{\Yv =\yv\} \cap\Ec_\text{aux})\leq 2\cdot 2^{-\frac{n}{2}\log\left(\frac{w_1^2 \sigma_a^2 + \sigma_z^2-\epsilon}{\sigma_z^2 + \epsilon^2 \sigma^2_a} \right)}\nonumber
\]
and therefore
\[
\sum_{s=2}^m \P(\Ec_s \cap \Ec_\text{aux})
\leq 2m \cdot 2^{-\frac{n}{2}\log\left(\frac{w_1^2 \sigma_a^2 + \sigma_z^2-\epsilon}{\sigma_z^2 + \epsilon^2 \sigma^2_a} \right)}\nonumber
\]
which tends to zero as $m \to \infty$, if
\begin{equation}
\limsup _{m\rightarrow \infty } \frac{\log m}{n_m} < \frac{1}{2}\log\left(\frac{w_1^2 \sigma_a^2 + \sigma_z^2-\epsilon}{\sigma_z^2 + \epsilon^2 \sigma^2_a} \right).
\label{case1_final}
\end{equation}
Therefore, by (\ref{C1_Error}), the probability of error averaged over
the random matrix $A^{(m)}$, $\P(\Ec)$, tends to zero as $m \to \infty$, if
(\ref{case1_final}) is satisfied, which in turn implies that there
exists a sequence of nonrandom measurement matrices
$\{A_0^{(m)}\}_{m=k}^\infty$ such that $\frac{1}{n_m m}\|A_0^{(m)}\|_F^2 \leq \sigma_a^2$ and
$\lim_{m\rightarrow\infty}\overline{P}_e(w_1, A_0^{(m)})=0, $ if
(\ref{case1_final}) is satisfied. Finally, since $\epsilon > 0$ is
chosen arbitrarily, we have the desired proof of Theorem~1.

\medskip

\textbf{Case 2}: $k \geq 2$. In this case, the signal of interest is $\Xv = \Xv(\wv, \Sv)$, where $\wv=[w_1, ..., w_k]^\intercal$ and $\Sv=[S_1,...,S_k]^\intercal$. Consider the following support recovery procedures. Fix $\epsilon > 0$. First, form an estimate $\hat{W}$ of $\|\mathbf{w}\|$ as
\begin{align}
\hat{W} \triangleq \sqrt{\frac{\left|\frac{1}{n}\|\Yv\|^2 - \sigma_z^2\right|}{\sigma_a^2}}.
\label{est_gain}
\end{align}

For $r,\zeta > 0$, let $\Qc=\Qc(r,\zeta)$ be a minimal set of points in $\mathbb{R}^k$ satisfying the following properties:
\begin{enumerate}
\item[i)] $\Qc \subseteq \mathcal{B}_k(r)$, where $\mathcal{B}_k(r)$ is the $k$-dimensional hypersphere of radius $r$.
\item[ii)] For any $\bv \in \Bc_k(r)$, there exists $\hat{\wv}\in\mathcal{Q}$ such that $\|\hat{\wv}-\bv\|\leq \frac{\zeta}{2}$.
\end{enumerate}

The following properties can be easily proved:
\begin{lemma}
\label{lemma2}
1) $\lim_{m\rightarrow\infty}\P\left(\exists \hat{\Wv}\in\Qc(\hat{W},\zeta)\textrm{ such that }\|\hat{\Wv}-\wv\| < \zeta\right) = 1.$

2) $q(r,\zeta)\triangleq|\Qc(r,\zeta)|$ is monotonically non-decreasing in $r$ for fixed $\zeta$.
\end{lemma}

\medskip

Given $\hat{W}$ and $\epsilon$, fix $\Qc = \Qc(\hat{W},\epsilon)$. Declare $d(\Yv) = \{\hat{s}_1, \hat{s}_2,...,\hat{s}_k\} \subseteq [m]$ is the recovered support of the signal, if it is the unique set of indices such that
\begin{align}
\frac{1}{n}\Bigg\|\Yv - \sum_{j=1}^k \hat{W}_{j} \Av_{\hat{s}_j}  \Bigg\|^2 \leq \sigma_z^2+\epsilon^2 \sigma_a^2
\label{C3_testing}
\end{align}
for some $\hat{\Wv}\in\mathcal{Q}$. If these is none or more than one such set, pick an arbitrary set of $k$ indices.

Next, we analyze the average probability of error
\[
\P(\Ec) = \E [\overline{P}_e(\wv, A)] = \P \{d^{(m)}(\Yv)\neq \{S_1, ..., S_k\}\}
\]
where the expectation is taken with respect to $A$. As before, we assume without loss of generality that $S_j = j$ for $j = 1, 2, ..., k$, which gives
\[
\Yv = \sum_{j=1}^k w_j\Av_j+\Zv\nonumber
\]
for some $\wv$. Define the event
\begin{align}
&\Ec_{s_1, s_2,...,s_k}\triangleq \nonumber\\
& \left\{\exists \hat{\Wv}\in \Qc \textrm{ and } \{s'_1, s'_2,...,s'_k\}= \{s_1, s_2,...,s_k\}\textrm{ such that }\frac{1}{n}\Bigg\|\Yv - \sum_{j=1}^k \hat{W}_{j} \Av_{s'_j} \Bigg\|^2\leq \sigma_z^2+\epsilon^2\sigma_a^2\right\}.\nonumber
\end{align}
Then
\begin{align}
\P(\Ec) &=  \P\left(\Ec_{1, 2,...,k}^c \cup \left(\bigcup_{s_1<\cdots<s_k:\{s_1, ..., s_k\}\neq[k]} \Ec_{s_1, s_2,...,s_k} \right)\right)\nonumber\\
&\leq  \P\left({\Ec^c_\text{aux}}\cup \Ec_{1, 2,...,k}^c   \cup\left(\bigcup_{s_1<\cdots<s_k:\{s_1, ..., s_k\}\neq[k]} (\Ec_{s_1, s_2,...,s_k}\cap\Ec_\text{aux}) \right)\right)\nonumber\\
&\leq  \P({\Ec^c_\text{aux}}) +
\P(\Ec_{1, 2,...,k}^c)+\sum_{s_1<\cdots<s_k:\{s_1, ..., s_k\}\neq[k]} \P(\Ec_{s_1, s_2,...,s_k}\cap\Ec_\text{aux})
\label{C2new_Perror}
\end{align}
where in this case
\begin{align}
\Ec_{\text{aux}}&\triangleq \left\{\hat{W}-\|\wv\|\in\left(-\epsilon,\epsilon\right)\right\}\cap
\left(\bigcap_{j=1}^k \left\{\frac{1}{n}\|\Av_j\|^2-\sigma_a^2 \in(-\epsilon, \epsilon)  \right\} \right)\nonumber\\
&~~~~ \cap \left(\bigcap_{j=1}^k\bigcap_{l= j+1}^k \left\{\frac{1}{n}\Av_j^\intercal\Av_l\in(-\epsilon, \epsilon)  \right\} \right)\cap\left( \bigcap_{j=1}^k \left\{\frac{1}{n}\Av_j^\intercal\Zv\in(-\epsilon, \epsilon)  \right\}\right)\nonumber\\
&~~~~\cap \left\{\frac{1}{n}\|\Zv\|^2-\sigma_z^2 \in(-\epsilon, \epsilon)  \right\}  .\nonumber
\end{align}

We now bound the terms in (\ref{C2new_Perror}). First, by the LLN, $\lim_{m\rightarrow \infty}\P(\Ec^c_\text{aux})=0$. Next, we consider $\P(\Ec_{1, 2,...,k}^c)$. Note that, for any $\hat{\Wv}\in\Qc$,
\begin{align}
\frac{1}{n}\Bigg\|\Yv - \sum_{j=1}^k \hat{W}_{j} \Av_{j} \Bigg\|^2 &=\frac{1}{n}\left\| \sum_{j=1}^k w_j\Av_j+\Zv-\sum_{j=1}^k \hat{W}_{j}\Av_j  \right\|^2\nonumber\\
&=\frac{1}{n} \sum_{j=1}^k\sum_{l=1}^k (w_j-\hat{W}_{j})(w_l-\hat{W}_{l})\Av_j^\intercal\Av_l +\frac{2}{n}\sum_{j=1}^k (w_j-\hat{W}_{j})\Av_j^\intercal\Zv+\frac{1}{n}\|\Zv\|^2.
\label{C2_Type1}
\end{align}
By applying the LLN to each term in (\ref{C2_Type1}), as similarly done in case 1, and using Lemma \ref{lemma2}-1), we have
\begin{align}
\lim_{m\rightarrow\infty}\P\left(\exists \hat{\Wv}\in\Qc\textrm{ such that  }\frac{1}{n}\Bigg\|\Yv - \sum_{j=1}^k \hat{W}_{j} \Av_{j} \Bigg\|^2   \leq \sigma_z^2+\epsilon^2\sigma_a^2 \right) = 1\nonumber
\end{align}
which implies that $
\lim_{m\rightarrow\infty}\P(\Ec_{1, 2,...,k}^c) = 0$.

Next, we consider $\P(\Ec_{s_1, s_2,...,s_k}\cap {\Ec_{\text{aux}}})$ for $\{s_1, s_2,...,s_k\} \neq [k]$. Note that
\begin{align}
&\P(\Ec_{s_1, s_2,...,s_k}\cap \Ec_{\text{aux}})\nonumber\\
&\leq \P(\Ec_{s_1, s_2,...,s_k}| \Ec_{\text{aux}})\nonumber\\
&= \int\cdots\int_{\{\av_1, ..., \av_k, \zv\}\in \Ec_{\text{aux}}}  \P(\Ec_{s_1, s_2,...,s_k} | \{\Av_1=\av_1\}\cap\cdots\cap\{\Av_k=\av_k\}\cap\{\Zv=\zv\}\cap\Ec_{\text{aux}})\nonumber\\
&~~~~~~~~~~~~~~~~~~~~~~~~~~~~~\times f(\av_1,...,\av_k,\zv|\Ec_{\text{aux}})d\av_1\cdots d\av_kd\zv.
\label{C3_oneside_prob}
\end{align}
For notational simplicity, define $\xi\triangleq\sigma_z^2+\epsilon^2 \sigma_a^2$, ${\Tc} \triangleq \{s_1, s_2,...,s_k\} \cap [k]$, ${\Tc}^c \triangleq \{s_1, s_2,...,s_k\}\backslash {\Tc}$, and $\Ec_{\textrm{cond}} \triangleq \{\Av_1=\av_1\}\cap\cdots\cap\{\Av_k=\av_k\}\cap\{\Zv=\zv\}\cap\Ec_{\text{aux}}$. For any permutation $(s_1',s_2', ..., s_k')$ of $\{s_1,s_2, ..., s_k\}$ and any $\hat{\Wv}\in\mathcal{Q}$,
\begin{align}
&\P\Bigg(\frac{1}{n}\Bigg\| \Yv-\sum_{j=1}^k \hat{W}_j \Av_{s_j'} \Bigg\|^2\leq\xi \bigg|\Ec_{\textrm{cond}}\Bigg)\nonumber\\
&=\P\Bigg(\frac{1}{n}\Bigg\| \sum_{j=1}^k w_j \Av_{j}+\Zv-\sum_{j=1}^k \hat{W}_j \Av_{s_j'} \Bigg\|^2\leq \xi\bigg|\Ec_{\textrm{cond}}\Bigg)\nonumber\\
&= \P\Bigg(\frac{1}{n}\Bigg\| \Bigg[\sum_{j=1}^k w_j \Av_{j}-\sum_{s_j'\in {\Tc}} \hat{W}_j \Av_{s_j'}+\Zv\Bigg]-\sum_{s_j' \in {\Tc}^c} \hat{W}_j \Av_{s_j'} \Bigg\|^2\leq \xi\bigg| \Ec_{\textrm{cond}}\Bigg).
\label{case2_lamma2}
\end{align}
Conditioned on $\Ec_{\textrm{cond}}$ and the chosen $\mathcal{Q}$, $\frac{1}{n}\big\|\sum_{j=1}^k w_j \Av_{j}-\sum_{s_j'\in {\Tc}} \hat{W}_j \Av_{s_j'}+\Zv \big\|^2$ is a fixed quantity satisfying
\begin{align}
\frac{1}{n}\Bigg\|\sum_{j=1}^k w_j \Av_{j}-\sum_{s_j'\in {\Tc}} \hat{W}_j \Av_{s_j'}+\Zv\Bigg\|^2&\in
 \Bigg(\Bigg[ \sum_{j\in[k]\backslash {\Tc}} w_j^2 + \sum_{s_j'\in {\Tc}}(w_{s_j'}-\hat{W}_j)^2 \Bigg]\sigma_a^2+\sigma_z^2 -\delta_1\epsilon,\nonumber\\
&~~~~~~\Bigg[ \sum_{j\in[k]\backslash {\Tc}} w_j^2 + \sum_{s_j'\in {\Tc}}(w_{s_j'}-\hat{W}_j)^2 \Bigg]\sigma_a^2+\sigma_z^2 +\delta_1\epsilon\Bigg)\nonumber
\end{align}
for some positive $\delta_1$ that depends on $\wv$ and $\epsilon$ only, and is non-decreasing in $\epsilon$. Meanwhile, $\Av_{s_j'}$ is independent of $\Av_1,...,\Av_k$, and $\Zv$ for $s_j'\in{\Tc}^c$. Hence, by Lemma \ref{lemma1} (with $\alpha = \big( \sum_{j\in[k]\backslash {\Tc}} w_j^2 + \sum_{s_j'\in {\Tc}}(w_{s_j'}-\hat{W}_j)^2 \big)\sigma_a^2+\sigma_z^2$ and $\gamma = \sigma_z^2+\epsilon^2\sigma_a^2$), (\ref{case2_lamma2}) is upper-bounded by
\begin{align}
2^{{-\frac{n}{2} \log \frac{\left( \sum\limits_{j\in[k]\backslash {\Tc}} w_j^2 + \sum\limits_{s_j'\in {\Tc}}\big(w_{s_j'}-\hat{W}_j\big)^2 \right)\sigma_a^2+\sigma_z^2 -\delta_1\epsilon} { \sigma_z^2+\epsilon^2\sigma_a^2  } }}\leq  2^{{-\frac{n}{2} \log \frac{\left( \sum\limits_{j\in[k]\backslash {\Tc}} w_j^2 \right)\sigma_a^2+\sigma_z^2 -\delta_1\epsilon} { \sigma_z^2+\epsilon^2\sigma_a^2  } }}.\nonumber
\end{align}
Hence, by the union of events bound,
\begin{align}
\P(\Ec_{s_1, s_2,...,s_k}| \Ec_{\textrm{cond}})
&\leq  \sum_{{\{s_1',...,s_k'\}=\{s_1,...,s_k\}}}  \P\Bigg(\exists \hat{\Wv}\in\Qc\textrm{ such that }\frac{1}{n}\Bigg\| \Yv-\sum_{j=1}^k \hat{W}_j \Av_{s_j'} \Bigg\|^2\leq \xi \bigg| \Ec_{\textrm{cond}}\Bigg)\nonumber\\
&\leq \sum_{{\{s_1',...,s_k'\}=\{s_1,...,s_k\}}} \sum_{\hat{\Wv}\in\Qc} \P\Bigg(\frac{1}{n}\Bigg\| \Yv-\sum_{j=1}^k \hat{W}_j \Av_{s_j'} \Bigg\|^2\leq \xi \bigg|  \Ec_{\textrm{cond}}\Bigg)\nonumber\\
&\leq k!\cdot|\mathcal{Q}|\cdot 2^ {- {\frac{ n }{2} \log \frac{\left( \sum\limits_{j\in[k]\backslash {\Tc}} w_j^2  \right)\sigma_a^2+\sigma_z^2 -\delta_1\epsilon} { \sigma_z^2+\epsilon^2 \sigma_a^2  } }}.\nonumber
\end{align}
Furthermore, conditioned on $\Ec_{\text{aux}}$, $\hat{W}<\|\wv\|+\epsilon$ and hence $|\mathcal{Q}| \leq q(\|\wv\|+\epsilon, \epsilon)$ by Lemma \ref{lemma2}-2). Thus,
\begin{align}
\P(\Ec_{s_1, s_2,...,s_k}\cap \Ec_{\text{aux}})\leq k! \cdot q(\|\wv\|+\epsilon, \epsilon) \cdot 2^ {- {\frac{ n }{2} \log \frac{\left( \sum\limits_{j\in[k]\backslash {\Tc}} w_j^2  \right)\sigma_a^2+\sigma_z^2 -\delta_1\epsilon} { \sigma_z^2+\epsilon^2 \sigma_a^2 } }}.
\label{general_bound}
\end{align}
Note that the probability upper-bound (\ref{general_bound}) depends on $s_1,...,s_k$ only through $\Tc$. Grouping the ${m-k}\choose {k-|{\Tc}|}$ events $\{\Ec_{s_1, s_2,...,s_k}\cap \Ec_{\text{aux}}\}$ with the same $\Tc$,
\begin{align}
\P(\Ec)
&\leq  \P(\Ec^c_\text{aux}) +\P(\Ec_{1, 2, ..., k}^c) +\sum_{{\Tc}\subset[k]} {{m-k}\choose {k-|{\Tc}|}} \cdot k!\cdot q(\|\wv\|+\epsilon, \epsilon) \cdot 2^{-\frac{n}{2} \log\frac{ \left(\sum\limits_{j\in[k]\backslash{\Tc}} w_j^2\right)  \sigma_a^2+\sigma_z^2 -\delta_1\epsilon} { \sigma_z^2+\epsilon^2\sigma_a^2  } }\nonumber\\
&\leq  \P(\Ec^c_\text{aux}) +\P(\Ec_{1, 2, ..., k}^c) +k! \cdot q(\|\wv\|+\epsilon, \epsilon) \cdot \sum_{{\Tc}\subset[k]} 2^{(k-|{\Tc}|)\log m}\cdot2^{-\frac{n}{2} \log\frac{ \left(\sum\limits_{j\in[k]\backslash{\Tc}} w_j^2  \right) \sigma_a^2+\sigma_z^2 -\delta_1\epsilon} { \sigma_z^2+\epsilon^2\sigma_a^2  } }\nonumber\\
&= \P(\Ec^c_\text{aux}) +\P(\Ec_{1, 2, ..., k}^c) +k! \cdot q(\|\wv\|+\epsilon, \epsilon)  \cdot\sum_{{\Tc}\subseteq[k]} 2^{|{\Tc}|\log m} \cdot 2^{-\frac{n}{2} \log\frac{ \left(\sum\limits_{j\in{\Tc}} w_j^2  \right) \sigma_a^2+\sigma_z^2 -\delta_1\epsilon} { \sigma_z^2+\epsilon^2\sigma_a^2  } }\nonumber
\end{align}
which tends to zero as $m\rightarrow \infty$, if
\begin{align}
\limsup_{m\rightarrow \infty}\frac{\log m}{n_m}<\frac{1}{2|{\Tc}|} \log\frac{ \left(\sum\limits_{j\in{\Tc}} w_j^2 \right) \sigma_a^2+\sigma_z^2 -\delta_1\epsilon} { \sigma_z^2+\epsilon^2 \sigma_a^2  }
\label{general_condition}
\end{align}
for all $\Tc\subseteq[k]$. Similar to the reasoning in case 1, it implies the existence of a sequence of nonrandom measurement matrices $\{A_0^{(m)}\}_{m=k}^\infty$ such that $\frac{1}{n_m m}\|A_0^{(m)}\|_F^2 \leq \sigma_a^2$ and $\lim_{m\rightarrow\infty}\overline{P}_e(\wv, A_0^{(m)})=0$ if (\ref{general_condition}) is satisfied. Since $\epsilon > 0$ is arbitrarily chosen, the proof of Theorem 1 is complete.

Now, it only remains to prove Lemma 1. For simplicity, let $\theta \equiv \sigma_v^2$. Denote $S_n=\frac{1}{n}\sum_{i=1}^n (u_i-V_i)^2$. The moment generating function of $S_n$ is
\begin{align}
\E[e^{tS_n}] = \E[e^{ \frac{t}{n}\sum_{i=1}^n (u_i-V_i)^2}] = \prod_{i=1}^{n} \E[e^{ \frac{t}{n} (u_i-V_i)^2}].
\label{charecteristic_function}
\end{align}
Note that $(u_i-V_i)^2/\theta$ is a noncentral $\chi^2$ random variable. Its moment generating function is given by \cite {Lancaster_x2} as $\E[e^{t(u_i-V_i)^2/\theta}] = \exp(\frac{t{u_i^2}/{\theta}}{1-2 t}) / (1-2t)^{\frac{1}{2}}$, for $t\leq 1/2$. By changing variable $\theta t/n\rightarrow t$, we have
\begin{align*}
\E[e^{t(u_i-V_i)^2/n}] &= \frac{e^{\frac{\frac{t}{n}{u_i^2}}{1-2 \theta t/n} }}{(1-2\theta t/n)^{\frac{1}{2}}}.
\end{align*}
Back to (\ref{charecteristic_function}), we obtain
\begin{align*}
\E[e^{tS_n}] = \prod_{i=1}^{n} \E[e^{ \frac{t}{n} (u_i-V_i)^2}] = \prod_{i=1}^{n}\frac{e^{\frac{\frac{t}{n}{u_i^2}}{1-2 \theta t/n} }}{(1-2\theta t/n)^{\frac{1}{2}}} = \frac{e^{\frac{\frac{t}{n}{ \sum_{i=1}^{n}u_i^2}}{1-2 \theta t/n} }}{(1-2\theta t/n)^{\frac{n}{2}}}.
\end{align*}

The Chernoff bound implies
\begin{align*}
\P(S_n \leq \gamma) &\leq  \min_{s>0}e^{s\gamma}\E[e^{-sS_n}]\\
&= \min_{s>0} e^{s\gamma} \frac{e^{\frac{-\frac{s}{n}{ \sum_{i=1}^{n}u_i^2}}{1+2 \theta s/n} }}{(1+2\theta s/n)^{\frac{n}{2}}}\\
&= \min_{p<0} e^{-p\gamma} \frac{e^{\frac{\frac{p}{n}{ \sum_{i=1}^{n}u_i^2}}{1-2 \theta p/n} }}{(1-2\theta p/n)^{\frac{n}{2}}}\\
&= \exp\left\{\min_{p<0}\left\{\log  e^{-p\gamma} \frac{e^{\frac{\frac{p}{n}{ \sum_{i=1}^{n}u_i^2}}{1-2 \theta p/n} }}{(1-2\theta p/n)^{\frac{n}{2}}} \right\} \right\}\\
&= \exp\left\{\min_{p<0}\left\{-p\gamma+ \frac{\frac{p}{n}{ \sum_{i=1}^{n}u_i^2}}{1-2 \theta p/n}  - {\frac{n}{2}}\log{(1-2\theta p/n)} \right\} \right\}.
\end{align*}
Define
\begin{align*}
f(p) &\triangleq -p\gamma+ \frac{\frac{p}{n}{ \sum_{i=1}^{n}u_i^2}}{1-2 \theta p/n}  - {\frac{n}{2}}\log{(1-2\theta p/n)}\\
g(\lambda) &\triangleq f(n\lambda) = -n\lambda\gamma+ \frac{\lambda{ \sum_{i=1}^{n}u_i^2}}{1-2 \theta \lambda}  - {\frac{n}{2}}\log{(1-2\theta \lambda)}.
\end{align*}

Clearly,
$
\min_{p<0}f(p) = \min_{\lambda<0}g(\lambda).
$
Denote
\begin{align*}
\alpha_s \triangleq \frac{1}{n} \sum_{i=1}^{n}u_i^2.
\end{align*}

Then, let us focus on the minimization problem
\begin{align*}
\min_{\lambda<0}g(\lambda)&= \min_{\lambda<0}\left\{ -n\lambda\gamma+ \frac{{n\lambda}\alpha_s}{1-2 \theta \lambda}  - {\frac{n}{2}}\log{(1-2\theta \lambda)}\right\}\\
&= n\cdot \min_{\lambda<0}\left\{ -\lambda\gamma+ \frac{\lambda\alpha_s}{1-2 \theta \lambda}  - {\frac{1}{2}}\log{(1-2\theta \lambda)}\right\}\\
&= -n\cdot \underbrace{\max_{\lambda<0}\left\{\lambda\gamma- \frac{\lambda\alpha_s}{1-2 \theta \lambda}  + {\frac{1}{2}}\log{(1-2\theta \lambda)}\right\}}_{\triangleq~\Lambda(\alpha_s, \theta, \gamma)}.
\end{align*}

It can be shown that the minimizing $\lambda$ is
\begin{align*}
\lambda^* = \frac{2\gamma-\theta-\sqrt{\theta^2 + 4\alpha_s\gamma}}{4\theta \gamma} <0
\end{align*}
and
\begin{align*}
\Lambda(\alpha_s, \theta, \gamma)
&=  \lambda^* \gamma - \frac{\lambda^* \alpha_s}{1-2\lambda^* \theta}+\frac{1}{2}\log(1-2\lambda^* \theta)\\
&= \frac{\alpha_s+\gamma}{2\theta} - \frac{1}{2}-\frac{2\alpha_s\gamma}{\theta(\theta+\sqrt{\theta^2+4\alpha_s\gamma})} +\frac{1}{2}\log \frac{\theta+\sqrt{\theta^2+4\alpha_s\gamma}}{2\gamma}.
\end{align*}
Next, for fixed $\alpha_s$ and $\gamma$,
\begin{align*}
\frac{\partial \Lambda(\alpha_s, \theta, \gamma)}{\partial \theta}
&= -\frac{\alpha_s+\gamma}{2\theta^2}+ \frac{2\alpha_s\gamma\left[\theta+\sqrt{\theta^2+4\alpha_s\gamma}+\theta\left(1+\frac{2\theta}{2\sqrt{\theta^2+4\alpha_s\gamma}}\right)\right]}{\theta^2(\theta+\sqrt{\theta^2+4\alpha_s\gamma})^2}\nonumber\\
&  ~~+ \frac{1}{2(\theta+\sqrt{\theta^2+4\alpha_s\gamma})}\left(1+\frac{2\theta}{2\sqrt{\theta^2+4\alpha_s\gamma}}\right)\nonumber\\
&= -\frac{\alpha_s+\gamma}{2\theta^2} + \frac{\sqrt{4\alpha_s\gamma+\theta^2}}{2\theta^2}.
\end{align*}
For $\theta>0$, there is only one stationary point $\theta' =\alpha_s-\gamma$, which is a solution to $\frac{\partial \Lambda(\alpha_s, \theta, \gamma)}{\partial \theta} = 0$. Check the second derivative,
\begin{align*}
\frac{\partial^2 \Lambda(\alpha_s, \theta, \gamma)}{\partial \theta^2}\Bigg|_{\theta = \alpha_s-\gamma} = \frac{1}{2(\alpha_s+\gamma)(\alpha_s-\gamma)} > 0.
\end{align*}
This confirms that $\theta' = \alpha_s-\gamma$ is the minimum point of $\Lambda(\alpha_s, \theta, \gamma)$, for $\theta>0$. Hence, for fixed $\alpha_s$ and $\gamma$ with $\gamma<\alpha_s$,
\begin{align*}
\Lambda(\alpha_s, \theta, \gamma) \geq \Lambda(\alpha_s, \theta', \gamma)
= \frac{1}{2}\log\frac{\alpha_s}{\gamma}.
\end{align*}
As a result,
\begin{align*}
\P(S_n \leq \gamma)
&\leq   \exp\left\{\min_{p<0}\left\{-p\gamma+ \frac{\frac{p}{n}{ \sum_{i=1}^{n}u_i^2}}{1-2 \theta p/n}  - {\frac{n}{2}}{(1-2\theta p/n)} \right\} \right\}\\
&= \exp\left\{\min_{\lambda<0}g(\lambda) \right\}\\
&= \exp\left\{-n\Lambda(\alpha_s, \theta, \gamma)\right\}\\
&\leq  \exp\left\{-n\Lambda(\alpha_s, \theta', \gamma)\right\}\\
&= \exp\left\{-\frac{n}{2}\log\left(\frac{\alpha_s}{\gamma}\right)\right\}\\
&\leq  \exp\left\{-\frac{n}{2}\log\left(\frac{\alpha-\beta}{\gamma}\right)\right\}.
\end{align*}

Hence, by changing the base of logarithm,
\begin{align*}
\P\left(\frac{1}{n}\sum_{i=1}^n(u_i-V_i)^2 \leq \gamma \right)
\leq 2^{-\frac{n}{2}\log\left(\frac{\alpha-\beta}{\gamma}\right)}.
\end{align*}

\section{Proof of Theorem 2}
\label{sec:Appen_Proof_Th2}
The main techniques for the proof of Theorem 2 include Fano's inequality and the properties of entropy. It mimics the proof of the converse for the channel coding theorem \cite{Cover06} with proper modification.

Assume there exist a sequence of measurement matrices $\{A^{(m)}\}_{m=k}^\infty$ and a sequence of support recovery maps $\{d^{(m)}\}_{m=k}^\infty$ such that
\begin{equation}
\frac{1}{n_m m}\|A^{(m)}\|_F^2 \leq \sigma_a^2
\label{Appen_proof_2.2_powerconstraint}
\end{equation}
and $\lim_{m\rightarrow\infty} \P(d^{(m)}(\Yv)\neq\{S_1, ..., S_k\}) =
0.$ We wish to show that
$\limsup_{m\rightarrow \infty} (\log m)/{n_m} \leq c(\wv).$

For any $\Tc\subseteq[k]$, denote the tuple of random
variables $(S_l:l\in\Tc)$ by $S(\Tc)$.  From Fano's inequality
\cite{Cover06}, we have
\begin{align}
H(S(\Tc)|\Yv) &\leq H(S_1, ..., S_k|\Yv) \nonumber\\
&\leq \log k!+ H(\{S_1, ..., S_k\}|\Yv)\nonumber\\
& \leq \log k! + \overline{P}_e(\wv,A^{(m)}) \log{m\choose k} + 1.
\label{T2_fanos}
\end{align}
For notation simplicity, let $\overline{P}_e^{(m)} \triangleq \overline{P}_e(\wv,A^{(m)})$. On the other hand,
\begin{align}
H(S(\Tc)|S(\Tc^c))
&=\log\left(\prod\limits_{q=0}^{|\Tc|-1}(m-(k-|\Tc|)-q)\right)\nonumber\\
&= |\Tc|\log m - n\epsilon_{1,n}
\label{general_starting_point}
\end{align}
where $\Tc^c \triangleq [k] \backslash \Tc$ and
\[
\epsilon_{1,n} \triangleq  \frac{1}{n} \log
\left({m^{|\Tc|}}/{\prod\limits_{q=0}^{|\Tc|-1}(m-(k-|\Tc|)-q)}\right)\nonumber
\]
which tends to zero as $n\rightarrow \infty$.  Hence, combining
(\ref{T2_fanos}) and (\ref{general_starting_point}), we have
\begin{align}
|\Tc|\log m&=H(S(\Tc)|S(\Tc^c))+n\epsilon_{1,n}\nonumber\\
&= I(S(\Tc);\Yv|S(\Tc^c)) + H(S(\Tc)|\Yv,S(\Tc^c))+n\epsilon_{1,n}\nonumber\\
&\leq  I(S(\Tc);\Yv|S(\Tc^c)) + H(S(\Tc)|\Yv)+n\epsilon_{1,n}\nonumber\\
&\leq  I(S(\Tc);\Yv|S(\Tc^c)) + \log k! + \overline{P}_e^{(m)} \log{m\choose k} +1+ n\epsilon_{1,n} \nonumber\\
&= \sum_{i=1}^n I(Y_i;S(\Tc)|Y_1^{i-1}, S(\Tc^c))+ \log k! + \overline{P}_e^{(m)} \log{m\choose k} + 1+ n\epsilon_{1,n}\nonumber\\
&\leq \sum_{i=1}^n \left(h(Y_i|S(\Tc^c))- h(Y_i|S_1,...,S_k)\right)+ \log k! + \overline{P}_e^{(m)} \log{m\choose k} + 1+ n\epsilon_{1,n}\nonumber\\
&= \sum_{i=1}^n (h(Y_i|S(\Tc^c))- h(Z_i))+ \log k! + \overline{P}_e^{(m)} \log{m\choose k} + 1+ n\epsilon_{1,n}\label{reduce_to_noise}
\end{align}
where (\ref{reduce_to_noise}) follows since the measurement matrix is fixed
and $Z_i$ is independent of $(S_1,\ldots, S_k)$.

Consider
\begin{align}
h(Y_i|S(\Tc^c)) &= h\left(\sum_{j=1}^k w_ja_{i, S_j}+Z_i \Big|S(\Tc^c)\right)  \nonumber\\
&= h\left(\sum_{j \in \Tc} w_j a_{i, S_j}+Z_i \Big|S(\Tc^c)\right)  \nonumber\\
&\leq  h\left(\sum_{j \in \Tc} w_ja_{i, S_j}+Z_i \right) \nonumber\\
&\leq  \frac{1}{2}\log\left(2\pi e\cdot
\textrm{Var}\left(\sum_{j \in \Tc} w_ja_{i, S_j} +Z_i \right)\right)
\label{property_different_entropy}
\end{align}
where the last inequality follows since the Gaussian random variable
maximizes the differential entropy given a variance constraint.
To further upper-bound (\ref{property_different_entropy}), note that
\[
{\textrm{Var}}\left(\sum_{j \in \Tc} w_ja_{i, S_j}+Z_i \right)=
\E\left[\left(\sum_{j \in \Tc} w_ja_{i, S_j}\right)^2\right]- \E\left[\sum_{j \in \Tc} w_ja_{i, S_j}\right]^2 + \sigma_z^2. \nonumber
\]
Now
\begin{align}
\E\left[\sum_{j \in \Tc} w_ja_{i, S_j}\right]
=\sum_{j \in \Tc} w_j\E[a_{i, S_j}]=\sum_{j \in \Tc} w_j \cdot \frac{1}{m}\sum_{p=1}^m a_{i, p}\nonumber
\end{align}
and
\begin{align}
\E\left[\left(\sum_{j \in \Tc} w_ja_{i, S_j}\right)^2\right]
&=\sum_{j \in \Tc} \sum_{ l\in \Tc} w_jw_l\E[a_{i, S_j}a_{i, S_l}]\nonumber\\
&=\sum_{j \in \Tc} \sum_{  l\in \Tc,l\neq j} w_jw_l\E[a_{i, S_j}a_{i, S_l}]+\sum_{j \in \Tc} w_j^2 \E[a_{i, S_j}^2]\nonumber\\
&=\sum_{j \in \Tc} \sum_{l \in \Tc,l\neq j} \frac{w_jw_l}{m(m-1)}\sum_{p=1}^m\sum_{\substack{q=1\\q\neq p}}^m a_{i,p}a_{i,q}+\sum_{j \in \Tc} w_j^2\cdot\frac{1}{m} \sum_{p=1}^m a_{i, p}^2\nonumber\\
&= \sum_{j \in \Tc} \sum_{l \in \Tc,l\neq j}
\frac{w_jw_l}{m(m-1)}\left(\sum_{p=1}^m a_{i,p}\right)^2
+\frac{1}{m}\left(\sum_{j \in \Tc} w_j^2 - \tau(m)\right)\sum_{p=1}^m a_{i, p}^2\nonumber
\end{align}
where $\tau(m)\triangleq \sum_{j \in \Tc} \sum_{l \in \Tc,l\neq j}
\frac{w_jw_l}{(m-1)} \to 0$ as $m \to \infty$.  It can be also easily checked that
\[
\sum_{j \in \Tc} \sum_{l \in \Tc,l\neq j} \frac{w_jw_l}{m(m-1)} \leq
\left(\sum_{j \in \Tc} \frac{w_j}{m}\right)^2 \nonumber
\]
and thus
\[
\textrm{Var}\left(\sum_{j \in \Tc} w_ja_{i, S_j} + Z_i\right)\leq
\left(\sum_{j \in \Tc} {w_j^2} - \tau(m)\right)\frac{1}{m}\sum_{p=1}^m
a_{i, p}^2 + \sigma_z^2.\nonumber
\]
Returning to (\ref{reduce_to_noise}), we have
\begin{align}
|\Tc|\log m
&\leq  \sum_{i=1}^n \frac{1}{2}\log\left[2\pi e \left( \Bigg(\sum_{j \in \Tc} {w_j^2} - \tau(m)\Bigg)\frac{1}{m}\sum_{p=1}^m a_{i, p}^2 +\sigma_z^2\right)\right] - \frac{n}{2}\log (2\pi e\sigma_z^2) \nonumber\\
&  ~~~~+ \log k! + \overline{P}_e^{(m)} \log{m\choose k} +1+ n\epsilon_{1,n}\nonumber\\
&\leq  \frac{n}{2}\log\left[2\pi e \left( \Bigg(\sum_{j \in \Tc} {w_j^2} - \tau(m)\Bigg)\frac{1}{nm}\sum_{i=1}^n \sum_{p=1}^m a_{i, p}^2 +\sigma_z^2\right)\right]  - \frac{n}{2}\log (2\pi e\sigma_z^2)  \nonumber\\
& ~~~~+ \log k! + \overline{P}_e^{(m)} \log{m\choose k} +1+ n\epsilon_{1,n}\label{Jensens}\\
&= \frac{n}{2}\log\left( \Bigg(\sum_{j \in \Tc} {w_j^2} - \tau(m)\Bigg)\frac{\sigma_a^2}{\sigma_z^2} +1\right)+ \log k! + \overline{P}_e^{(m)} \log{m\choose k} + 1+n\epsilon_{1,n}
\label{use_power_constraint}
\end{align}
where (\ref{Jensens}) is due to Jensen's inequality and
(\ref{use_power_constraint}) follows from
(\ref{Appen_proof_2.2_powerconstraint}). Therefore,
\[
\limsup_{m\rightarrow\infty}\frac{\log m}{n_m} - \frac{\log k! + \overline{P}_e^{(m)} \log{m\choose k} + 1+ n_m\epsilon_{1,n_m}}{|\Tc|n_m} \leq \frac{1}{2|\Tc|}\log \left( 1+\frac{\sigma_a^2}{\sigma_z^2} \sum_{j\in \Tc} w_j^2  \right)
\]
for all $\Tc\subseteq [k]$. Due to the fact that $\log{m\choose k} \leq k\log m$, we have
\[
\limsup_{m\rightarrow\infty}\frac{(1-k\overline{P}_e^{(m)}/|\Tc|)\log m}{n_m} - \frac{\log k! + n_m\epsilon_{1,n_m} + 1}{|\Tc|n_m} \leq \frac{1}{2|\Tc|}\log \left( 1+\frac{\sigma_a^2}{\sigma_z^2} \sum_{j\in \Tc} w_j^2  \right)
\]
for all $\Tc\subseteq [k]$. Since $\lim_{m\rightarrow\infty} \overline{P}_e^{(m)} = 0$, we reach the conclusion
\[
\limsup_{m\rightarrow\infty}\frac{\log m}{n_m} \leq \frac{1}{2|\Tc|}\log \left( 1+\frac{\sigma_a^2}{\sigma_z^2} \sum_{j\in \Tc} w_j^2  \right)
\]
for all $\Tc\subseteq [k]$, which completes the proof of Theorem 2.

\section{Proof of Theorem 3}
\label{sec:ProofTh3}
We show that
\[
\lim_{m\rightarrow \infty} \P\{d^{(m)}(A\Xv(\wv^{(m)}, \Sv) + \Zv)\neq \textmd{supp}(\Xv(\wv^{(m)}, \Sv))\}=0\nonumber
\]
provided that the condition
\begin{align}
\limsup_{m\rightarrow \infty}  \frac{1}{n_m} \max_{j\in[k_m]}\left[\frac{6k_m\log k_m + 2j\log m}{\log\left(\frac{j w_{\text{min}}^2\sigma_a^2}{\sigma_z^2}+1\right)}\right] < 1
\label{repeat_condition_T3}
\end{align}
is satisfied. Note that (\ref{repeat_condition_T3}) implies that $n = \max[\Omega(k\log k),\Omega ({\frac{k}{\log k}\log m})]$, which in turn implies that $k=o(n)$.

We follow the proof of Theorem 1 in Appendix \ref{sec4}. Recall that in case 2 of the proof of Theorem 1, we first proposed the support recovery rule (\ref{C3_testing}). Then, we formed estimates of the nonzero values, and used them to test all possible sets of $k$ indices. The key step was to analyze two types of errors. On the one hand, the true support should satisfy the reconstruction rule (\ref{C3_testing}) with high probability. On the other hand, the probability that at least one incorrect support possibility satisfies this rule was controlled to diminish as the problem size increases.

By mainly replicating the steps in Appendix \ref{sec4} with necessary accommodations to the new setting with growing number of nonzero entries, we present the proof of Theorem 3 as follows.

\begin{enumerate}
\item We first modify the support recovery rule by replacing (\ref{C3_testing}) with
\begin{align}
\frac{1}{n}\Bigg\|\Yv - \sum_{j=1}^k \hat{W}_{j} \Av_{\hat{s}_j}  \Bigg\|^2 \leq (1+\epsilon)\sigma_z^2+2\epsilon^2 \sigma_a^2.
\label{New_extend_rule}
\end{align}

\item The cardinality $q(r,\zeta)$ of a minimal $\Qc(r,\zeta)$ can be upper-bounded by
\[
q(r,\zeta) \leq \left(\frac{\eta_1 kr}{\zeta}\right)^{k}\nonumber
\]
for some $\eta_1>0$. This can be easily shown by first partitioning the $k$-dimensional hypercube of side $2r$ into identical elementary hypercubes with side not exceeding $\frac{\zeta}{4k}$ and then, for each elementary hypercube that intersects the hypersphere, picking an arbitrary point on the hypersphere within that elementary hypercube. The resulting set of points provides the upper bound above for $q(r,\zeta)$.

\item Define $\sigma_{\textrm{max}}^2$ and $\sigma_{\textrm{min}}^2$ to be the largest and smallest eigenvalues of the matrix
    \[\frac{1}{{n\sigma_a^2}}[\Av_1,...,\Av_k, \frac{\sigma_a}{\sigma_z}\Zv]^\intercal[\Av_1,...,\Av_k,\frac{\sigma_a}{\sigma_z}\Zv]\]
     respectively. We replace the definition of $\Ec_{\textrm{aux}}$ by
\[
\Ec_{\text{aux}}\triangleq \left\{\hat{W}-\|\wv\|\in\left(-\epsilon,\epsilon\right)\right\}\cap
\left\{ \sigma_{\textrm{max}}^2 \in \left(1 - \epsilon, 1 + \epsilon\right) \right\} \cap \left\{ \sigma_{\textrm{min}}^2 \in \left(1 - \epsilon, 1 + \epsilon\right) \right\}.\nonumber
\]

Consider the asymptotic behaviors of the events. First, note that
\begin{align}
\sqrt{\frac{1}{n\sigma_a^2}\|\Yv\|^2} = \sqrt{\frac{\|\wv\|^2\sigma_a^2+\sigma_z^2}{n\sigma_a^2}}\sqrt{\left\|\frac{\Yv}{\sqrt{\|\wv\|^2\sigma_a^2+\sigma_z^2}}\right\|^2}
\end{align}
where $\sqrt{\left\|\frac{\Yv}{\sqrt{\|\wv\|^2\sigma_a^2+\sigma_z^2}}\right\|^2}$ is $\chi$-distributed with mean $\sqrt{2}\,\frac{\Gamma((n+1)/2)}{\Gamma(n/2)}$ and variance $\left(n-\frac{2\Gamma^2((n+1)/2)}{\Gamma^2(n/2)}\right)$. Then, $\sqrt{\frac{1}{n\sigma_a^2}\|\Yv\|^2}$ has mean  $\sqrt{\frac{\|\wv\|^2\sigma_a^2+\sigma_z^2}{n\sigma_a^2}}\sqrt{2}\,\frac{\Gamma((n+1)/2)}{\Gamma(n/2)}$ and variance $\frac{\|\wv\|^2\sigma_a^2+\sigma_z^2}{n\sigma_a^2}\left(n-\frac{2\Gamma^2((n+1)/2)}{\Gamma^2(n/2)} \right)$.

It has been shown \cite{CHENQI} that
\[
\lim_{x\rightarrow\infty}\frac{x\Gamma(x)}{\sqrt{x+1/4}\Gamma(x+1/2)} = 1.
\]
Then, as $n\rightarrow\infty$, $\sqrt{\frac{1}{n\sigma_a^2}\|\Yv\|^2}$ has asymptotic mean $\sqrt{\frac{\|\wv\|^2\sigma_a^2+\sigma_z^2}{\sigma_a^2}}$ and variance $\frac{\|\wv\|^2\sigma_a^2+\sigma_z^2}{2n\sigma_a^2}$. Since $k=o(n)$, we have $\frac{\|\wv\|^2\sigma_a^2+\sigma_z^2}{2n\sigma_a^2}\rightarrow 0$. Hence, $\lim_{m\rightarrow\infty}\P\{\hat{W}-\|\wv\|\in\left(-\epsilon,\epsilon\right)\}=1$.

Second, $\sigma_{\textrm{max}}^2$ and $\sigma_{\textrm{min}}^2$ are shown \cite{Silverstein} to almost surely converge to $(1+q)^2$ and $(1-q)^2$, respectively, where $q\triangleq\lim_{m\rightarrow\infty}\sqrt{(k+1)/n}=0$. Thus, $\lim_{m\rightarrow\infty} \P(\Ec_{\text{aux}}^c) = 0$. 

\item Next, we analyze the probability that the true support satisfies the recovery rule. Note that
\begin{align}
\frac{1}{n}\Bigg\|\Yv - \sum_{j=1}^k \hat{W}_{j} \Av_{j} \Bigg\|^2 &= \frac{1}{n}\left\| \sum_{j=1}^k w_j\Av_j+\Zv-\sum_{j=1}^k \hat{W}_{j}\Av_j  \right\|^2\nonumber\\
&= \frac{1}{n}\left\| \left[\Av_1,...,\Av_k, \frac{\sigma_a}{\sigma_z}\Zv\right]\left[\begin{array}{c} \wv-\hat{\Wv}\\\frac{\sigma_z}{\sigma_a}\end{array}\right]\right\|^2\nonumber\\
&\leq \sigma_{\textrm{max}}^2 \sigma_a^2 \left\|\left[\begin{array}{c} \wv-\hat{\Wv}\\\frac{\sigma_z}{\sigma_a}\end{array}\right]\right\|^2\nonumber\\
&=  \sigma_{\textrm{max}}^2 \sigma_a^2 \|\wv-\hat{\Wv}\|^2 + \sigma_{\textrm{max}}^2 \sigma_z^2.
\label{General_growK_Type1}
\end{align}
By using the fact that $\sigma_{\textrm{max}}^2\rightarrow 1$ almost surely as $n\rightarrow \infty$ and Lemma 2-1), we have $\lim_{m\rightarrow\infty}\P(\Ec_{1, 2,...,k}^c) = 0$.

\item Now, suppose we have proceeded to a step similar to (\ref{case2_lamma2}) (that is, to be exact, equipped with the modified rule (\ref{New_extend_rule}) and a proper $\Ec_{\text{cond}}$). Define the auxiliary vector $\wv'\in{\mathbb{R}^{k+1}}$ as
\begin{align}
w'_j = \left\{ \begin{array}{ll}
w_j & \mbox{if $j\in[k]\backslash \Tc$},\\
w_j - \hat{W}_i & \mbox{if $j=s_i' \in \Tc$},\\
\frac{\sigma_z}{\sigma_a} &  \mbox{if $j=k+1$}.\end{array} \right.
\end{align}
Then,
\begin{align}
\frac{1}{n}\left\|\sum_{j=1}^k w_j \Av_{j}-\sum_{s_j'\in {\Tc}} \hat{W}_j \Av_{s_j'}+\Zv\right\|^2
&=\frac{1}{n}\left\| \left[\Av_1,...,\Av_k, \frac{\sigma_a}{\sigma_z}\Zv\right]\wv'\right\|^2\nonumber\\
&\geq \left(1-\epsilon\right)\|\wv'\|^2\sigma_a^2\nonumber\\
&\geq \left(1-\epsilon\right)\left(\left(\sum\limits_{j\in[k]\backslash \Tc} w_j^2\right)\sigma_a^2 + \sigma_z^2\right).\nonumber
\end{align}
From Lemma 1, it follows that (for sufficiently small $\epsilon$)
\begin{align}
\P\left(\frac{1}{n}\Bigg\| \Yv-\sum_{j=1}^k \hat{W}_j \Av_{s_j'} \Bigg\|^2\leq(1+\epsilon)\sigma_z^2+2\epsilon^2 \sigma_a^2 \bigg|\Ec_{\textrm{cond}}\right) \leq 2^{{-\frac{n}{2} \log \frac{\left(1-\epsilon\right)\left(\left(\sum\limits_{j\in[k]\backslash \Tc} w_j^2\right)\sigma_a^2 + \sigma_z^2\right)} { (1+\epsilon)\sigma_z^2+2\epsilon^2 \sigma_a^2 } }}.\nonumber
\end{align}

\item With these modifications, we follow the proof steps of Theorem 1 to reach
\begin{align}
\P(\Ec)
&\leq \P(\Ec^c_\text{aux}) +\P(\Ec_{1, 2, ..., k}^c) +k! \cdot q(\|\wv\|+\epsilon, \epsilon)  \cdot\sum_{{\Tc}\subseteq[k]} 2^{|{\Tc}|\log m} \cdot 2^{{-\frac{n}{2} \log \frac{\left(1-\epsilon\right)\left(\left(\sum\limits_{j\in \Tc} w_j^2\right)\sigma_a^2 + \sigma_z^2\right)} {(1+\epsilon)\sigma_z^2+2\epsilon^2 \sigma_a^2  } }}\nonumber\\
&\leq \P(\Ec^c_\text{aux}) +\P(\Ec_{1, 2, ..., k}^c) +k! \cdot q(\|\wv\|+\epsilon, \epsilon)  \cdot\sum_{{\Tc}\subseteq[k]} 2^{|{\Tc}|\log m} \cdot 2^{{-\frac{n}{2} \log \frac{\left(1-\epsilon\right)\left(|{\Tc}| w_{\textrm{min}}^2\sigma_a^2 + \sigma_z^2\right)} {(1+\epsilon)\sigma_z^2+2\epsilon^2 \sigma_a^2  } }}\nonumber\\
&\leq \P(\Ec^c_\text{aux}) +\P(\Ec_{1, 2, ..., k}^c) +k! \cdot q(\|\wv\|+\epsilon, \epsilon)  \cdot 2^k \cdot \max_{j\in[k]} \left[2^{j\log m} \cdot 2^{{-\frac{n}{2} \log \frac{\left(1-\epsilon\right)\left(j w_{\textrm{min}}^2\sigma_a^2 + \sigma_z^2\right)} {(1+\epsilon)\sigma_z^2+2\epsilon^2 \sigma_a^2  } }}\right].
\label{growingK_overall_bound}
\end{align}

Note that
\begin{align}
&\log \left(k! \cdot q(\|\wv\|+\epsilon, \epsilon) \cdot   2^k \cdot \max_{j\in[k]} \left[2^{j\log m} \cdot 2^{{-\frac{n}{2} \log \frac{\left(1-\epsilon\right)\left(j w_{\textrm{min}}^2\sigma_a^2 + \sigma_z^2\right)} {(1+\epsilon)\sigma_z^2+2\epsilon^2 \sigma_a^2  } }}\right] \right)\nonumber\\
&\leq k\log k + k\log (\eta_1 k^2 w_{\textrm{max}}/\epsilon) + k + \max_{j\in[k]} \left[j\log m -\frac{n}{2} \log\frac{\left(1-\epsilon\right)\left(j w_{\textrm{min}}^2\sigma_a^2 + \sigma_z^2\right)} { (1+\epsilon)\sigma_z^2+2\epsilon^2 \sigma_a^2 }\right].
\label{Grow_K_bound}
\end{align}
It can be readily seen that from the condition (\ref{repeat_condition_T3}), the upper bound in (\ref{Grow_K_bound}) becomes negative and thus $\P(\Ec)\rightarrow 0$ as $m\rightarrow \infty$.
\end{enumerate}

\section{Proof of Theorem 4}
\label{sec:Proof_Th4}
To establish this theorem, we prove the following equivalent statement:

If there exists a sequence of matrices $\{A^{(m)}\}_{m=k}^\infty$, $A^{(m)}\in\mathbb{R}^{n_m\times m}$, and a sequence of support recovery maps $\{d^{(m)}\}_{m=k}^\infty$, $d^{(m)}:\mathbb{R}^{n_m}\mapsto 2^{\{1, 2, ..., m\}}$, such that
\[
\frac{1}{n_m m}\|A^{(m)}\|_F^2 \leq \sigma_a^2
\]
and
\[
\lim_{m\rightarrow\infty} \overline{P}_e(\wv^{(m)},A^{(m)}) = 0
\]
then
\[
\limsup_{m\rightarrow\infty}\frac{2k_m \log ({m}/{k_m}) }{n_m\log\left( \frac{2k_m w_{\textrm{max}}^2\sigma_a^2}{\sigma_z^2} +1\right)} \leq 1.
\]

To justify this alternative claim, we follow the steps for the proof of Theorem 2 in Appendix \ref{sec:Appen_Proof_Th2}. Necessary modifications and clarifications are presented as follows.
\begin{enumerate}
\item Note that
\begin{align}
\epsilon_{1,n} =  \frac{1}{n} \log
\left({m^{|\Tc|}}/{\prod\limits_{q=0}^{|\Tc|-1}\left(m-(k-|\Tc|)-q\right)}\right) \leq \frac{|\Tc|}{n} \log\frac{m}{m-k+1}.
\label{bound_epsilon}
\end{align}


\item For any $\Tc\subseteq[k]$, we follow (\ref{use_power_constraint}) in Appendix \ref{sec:Appen_Proof_Th2} to reach
\begin{align}
|\Tc|\log m - \log k! - \overline{P}_e^{(m)} \log{m\choose k} -1- n\epsilon_{1,n} \leq \frac{n}{2}\log\left( \Bigg(\sum_{j \in \Tc} {w_j^2} - \tau(m)\Bigg)\frac{\sigma_a^2}{\sigma_z^2} +1\right).
\label{General_converse_1}
\end{align}
Note that
\begin{align}
\sum_{j \in \Tc} {w_j^2} - \tau(m) &\leq \left|\sum_{j \in \Tc} {w_j^2}\right| + \left| \sum_{j \in \Tc} \sum_{l \in \Tc,l\neq j} \frac{w_jw_l}{(m-1)}\right|\nonumber\\
& \leq |\Tc| w_{\textrm{max}}^2 +  \frac{|\Tc|(|\Tc|-1)w_{\textrm{max}}^2}{(m-1)}\nonumber\\
& \leq 2|\Tc|w_{\textrm{max}}^2.
\label{General_Converse_2}
\end{align}
Then, it follows from (\ref{bound_epsilon}), (\ref{General_converse_1}), and (\ref{General_Converse_2}) that the inequality
\[
|\Tc|\log m - k\log k - \overline{P}_e^{(m)} k\log m -1- n\cdot\frac{|\Tc|}{n} \log\frac{m}{m-k+1} \leq \frac{n}{2}\log\left( 2|\Tc|w_{\textrm{max}}^2\frac{\sigma_a^2}{\sigma_z^2} +1\right)\nonumber
\]
must hold for any $\Tc\subseteq[k]$. By choosing $\Tc = [k]$, we have
\begin{align}
\limsup_{m\rightarrow\infty}\frac{2k_m\left(\log (m-k_m+1) - \log k_m - \frac{1}{k_m}\right)}{n_m\log\left( \frac{2k_m w_{\textrm{max}}^2\sigma_a^2}{\sigma_z^2} +1\right)} \leq 1 \nonumber
\end{align}
which equivalently gives
\begin{align}
\limsup_{m\rightarrow\infty}\frac{2k_m\left(\log m - \log k_m \right)}{n_m\log\left( \frac{2k_m w_{\textrm{max}}^2\sigma_a^2}{\sigma_z^2} +1\right)} \leq 1. \nonumber
\end{align}
\end{enumerate}


\begin{thebibliography}{10}
\providecommand{\url}[1]{#1}
\csname url@samestyle\endcsname
\providecommand{\newblock}{\relax}
\providecommand{\bibinfo}[2]{#2}
\providecommand{\BIBentrySTDinterwordspacing}{\spaceskip=0pt\relax}
\providecommand{\BIBentryALTinterwordstretchfactor}{4}
\providecommand{\BIBentryALTinterwordspacing}{\spaceskip=\fontdimen2\font plus
\BIBentryALTinterwordstretchfactor\fontdimen3\font minus
  \fontdimen4\font\relax}
\providecommand{\BIBforeignlanguage}[2]{{%
\expandafter\ifx\csname l@#1\endcsname\relax
\typeout{** WARNING: IEEEtran.bst: No hyphenation pattern has been}%
\typeout{** loaded for the language `#1'. Using the pattern for}%
\typeout{** the default language instead.}%
\else
\language=\csname l@#1\endcsname
\fi
#2}}
\providecommand{\BIBdecl}{\relax}
\BIBdecl

\bibitem{Donoho06_4}
D.~L. Donoho, ``Compressed sensing,'' \emph{IEEE Trans. Inform. Theory},
  vol.~52, no.~4, pp. 1289--1306, 2006.

\bibitem{Candes_comp}
E.~J. Candes, ``Compressive sampling,'' \emph{Proceedings of the Int. Congress
  of Mathematicians}, pp. 1433--1452, 2006.

\bibitem{Rao97}
I.~Gorodnitsky and B.~Rao, ``Sparse signal reconstruction from limited data
  using focuss: A re-weighted norm minimization algorithm,'' \emph{IEEE Trans.
  Sig. Proc.}, vol.~45, no.~3, pp. 600--616, 1997.

\bibitem{Rao98}
I.~F. Gorodnitsky, J.~S. George, and B.~D. Rao, ``"neuromagnetic source imaging
  with focuss: A recursive weighted minimum norm algorithm,'' \emph{J.
  Electroencephalog. Clinical Neurophysiol.}, vol.~95, pp. 231–--251, 1995.

\bibitem{Jeffs98}
B.~D. Jeffs, ``Sparse inverse solution methods for signal and image processing
  applications,'' \emph{Proc. ICASSP}, pp. 1885–--1888, 1998.

\bibitem{Baraniuksinglepixel}
M.~Duarte, M.~Davenport, D.~Takhar, J.~Laska, T.~Sun, K.~Kelly, and R.~G.
  Baraniuk, ``Single-pixel imaging via compressive sampling,'' \emph{IEEE
  Signal Processing Magazine}, vol.~25, pp. 83--91, 2008.

\bibitem{Parks91}
S.~D. Cabrera and T.~W. Parks, ``Extrapolation and spectral estimation with
  iterative weighted norm modification,'' \emph{IEEE Trans. Acoust., Speech,
  Signal Process.}, vol.~4, pp. 842–--851, 1991.

\bibitem{JR_10ICASSP}
Y.~Jin and B.~D. Rao, ``Algorithms for robust linear regression by exploiting
  the connection to sparse signal recovery,'' \emph{ICASSP}, 2010.

\bibitem{ChuSpeech}
W.~C. Chu, \emph{Speech coding algorithms}.\hskip 1em plus 0.5em minus
  0.4em\relax Wiley-Interscience, 2003.

\bibitem{RaoCotter02}
S.~F. Cotter and B.~D. Rao, ``Sparse channel estimation via matching pursuit
  with application to equalization,'' \emph{IEEE Trans. on Communications},
  vol.~50, pp. 374--377, 2002.

\bibitem{NowakChannel}
W.~U. Bajwa, J.~Haupt, G.~Raz, and R.~Nowak, ``Compressed channel sensing,''
  \emph{Proc. of CISS}, 2008.

\bibitem{Duttweiler2000}
D.~L. Duttweiler, ``Proportionate normalized least-mean-squares adaptation in
  echo cancelers,'' \emph{IEEE Trans. Acoust., Speech, Signal Process.},
  vol.~8, pp. 508–--518, 2000.

\bibitem{RaoSong}
B.~D. Rao and B.~Song, ``Adaptive filtering algorithms for promoting
  sparsity,'' \emph{ICASSP}, pp. 361--364, 2003.

\bibitem{Guo09ITA}
D.~Guo, ``Neighbor discovery in ad hoc networks as a compressed sensing
  problem,'' \emph{Presented at Information Theory and Application Workshop,
  UCSD}, 2009.

\bibitem{Zhang93}
S.~G. Mallat and Z.~Zhang, ``Matching pursuits with time-frequency
  dictionaries,'' \emph{IEEE Trans. Sig. Proc.}, vol.~41, no.~12, pp.
  3397--3415, 1993.

\bibitem{Pati93}
Y.~C. Pati, R.~Rezaiifar, and P.~S. Krishnaprasad, ``Orthogonal matching
  pursuit: Recursive function approximation with applications to wavelet
  decomposition,'' \emph{27th Annual Asilomar Conf. Sig. Systems. Computers.},
  1993.

\bibitem{Tibshirani94}
R.~Tibshirani, ``Regression shrinkage and selection via the lasso,'' \emph{J.
  R. Statist. Soc. B}, vol.~58, no.~1, pp. 267--288, 1996.

\bibitem{Donoho01}
S.~S. Chen, D.~L. Donoho, and M.~A. Saunders, ``Atomic decomposition by basis
  pursuit,'' \emph{SIREV}, vol.~43, no.~1, pp. 129--159, 2001.

\bibitem{Tipping01}
M.~E. Tipping, ``Sparse bayesian learning and the relevance vector machine,''
  \emph{JMLR}, 2001.

\bibitem{Vetterli02}
M.~Vetterli, P.~Marziliano, and T.~Blu, ``Sampling signals with finite rate of
  innovation,'' \emph{IEEE Trans. on Signal Processing}, vol.~50, pp.
  1417--1428, 2002.

\bibitem{NeedellTropp08}
D.~Needell and J.~A. Tropp, ``Cosamp: Iterative signal recovery from incomplete
  and inaccurate samples,'' \emph{preprint}, 2008.

\bibitem{DaiMilenkovic08}
W.~Dai and O.~Milenkovic, ``Subspace pursuit for compressive sensing: Closing
  the gap between performance and complexity,'' \emph{preprint}, 2008.

\bibitem{Donoho06_1}
D.~Donoho, M.~Elad, and V.~N. Temlyakov, ``Stable recovery of sparse
  overcomplete representations in the presense of noise,'' \emph{IEEE Trans.
  Inform. Theory}, vol.~52, no.~1, pp. 6--18, 2006.

\bibitem{Tao05a}
E.~J. Candes and T.~Tao, ``Decoding by linear programming,'' \emph{IEEE Trans.
  Inform. Theory}, vol.~51, no.~12, pp. 4203--4215, 2005.

\bibitem{Tao06}
E.~J. Candes, J.~K. Romberg, and T.~Tao, ``Stable signal recovery from
  incomplete and inaccurate measurements,'' \emph{Comm. Pure Appl. Math}, 2006.

\bibitem{Tropp04}
J.~A. Tropp, ``Greedy is good: Algorithmic results for sparse approximation,''
  \emph{IEEE Transactions on Information Theory}, vol.~50, no.~10, pp.
  2231--2242, 2004.

\bibitem{Tropp07}
J.~A. Tropp and A.~C. Gilbert, ``Signal recovery from random measurements via
  orthogonal matching pursuit,'' \emph{IEEE Transactions on Information
  Theory}, vol.~53, no.~12, pp. 4655--4666, 2007.

\bibitem{Donoho06}
D.~Donoho, Y.~Tsaig, I.~Drori, and J.~Starck, ``Sparse solution of
  underdetermined linear equations by stagewise orthogonal matching pursuit,''
  \emph{preprint}, 2006.

\bibitem{Huo01}
D.~L. Donoho and X.~Huo, ``Uncertainty principles and ideal atomic
  decomposition,'' \emph{IEEE Trans. Info. Theory}, vol.~47, pp. 2845--2862,
  2001.

\bibitem{Leahy_MEGEEG}
S.~Baillet, J.~C. Mosher, and R.~M. Leahy, ``Electromagnetic brain mapping,''
  \emph{IEEE Signal Process. Mag.}, pp. 14–--30, 2001.

\bibitem{WipfNaga2008}
D.~Wipf and S.~Nagarajan, ``A unified bayesian framework for meg/eeg source
  imaging,'' \emph{NeuroImage}, pp. 947–--966, 2008.

\bibitem{Giannakis2007ICASSP}
Z.~Tian and G.~B. Giannakis, ``Compressed sensing for wideband cognitive
  radios,'' \emph{ICASSP}, pp. 1357–--1360, 2007.

\bibitem{Wain07}
M.~Wainwright, ``Information-theoretic bounds on sparsity recovery in the
  high-dimensional and noisy setting,'' \emph{Proc. Int'l. Symp. Info. Theo.},
  June 2007.

\bibitem{Wain09b}
------, ``Information-theoretic limits on sparsity recovery in the
  high-dimensional and noisy setting,'' \emph{IEEE Transactions on Information
  Theory}, no.~12, pp. 5728--5741, Dec 2009.

\bibitem{Fletcher08NIPS}
A.~K. Fletcher, S.~Rangan, and V.~K. Goyal, ``Resolution limits of sparse
  coding in high dimensions,'' \emph{NIPS}, 2008.

\bibitem{Fletcher09IT}
------, ``Necessary and sufficient conditions for sparsity pattern recovery,''
  \emph{IEEE Transactions on Information Theory}, vol.~55, no.~12, pp.
  5758--5772, Dec 2009.

\bibitem{Wang08ISIT}
W.~Wang, M.~J. Wainwright, and K.~Ramchandran, ``Information-theoretic limits
  on sparse signal recovery: Dense versus sparse measurement,'' \emph{Proc. of
  ISIT}, pp. 2197--2201, 2008.

\bibitem{Akcakaya}
M.~Ak\c{c}akaya and V.~Tarokh, ``Shannon theoretic limits on noisy compressive
  sampling,'' \emph{IEEE Transactions on Information Theory}, vol.~56, no.~1,
  pp. 492--504, 2010.

\bibitem{Bara06}
S.~Sarvotham, D.~Baron, and R.~G. Baraniuk, ``Measurements vs. bits: Compressed
  sensing meets information theory,'' \emph{Proceedings of 44th Allerton Conf.
  Comm., Ctrl., Computing}, 2006.

\bibitem{Fletcher_07}
A.~K. Fletcher, S.~Rangan, and V.~K. Goyal, ``On the rate-distortion
  performance of compressed sensing,'' \emph{ICASSP}, pp. 885--888, April 2007.

\bibitem{Hassibi09}
S.~Jafarpour, W.~Xu, B.~Hassibi, and R.~Calderbank, ``Efficient and robust
  compressed sensing using optimized expander graphs,'' \emph{to appear in IEEE
  Trans. Info. Theory}, 2009.

\bibitem{Milenkovic09}
H.~V. Pham, W.~Dai, and O.~Milenkovic, ``Sublinear compressive sensing
  reconstruction via belief propagation decoding,'' \emph{Preprint}, 2009.

\bibitem{Pfister08}
F.~Zhang and H.~D. Pfister, ``Compressed sensing and linear codes over real
  numbers,'' \emph{Information Theory and Applications Workshop, UCSD}, 2008.

\bibitem{Jin08}
Y.~Jin and B.~D. Rao, ``Insights into the stable recovery of sparse solutions
  in overcomplete representations using network information theory,''
  \emph{ICASSP}, pp. 3921 -- 3924, 2008.

\bibitem{Jin_ISIT}
------, ``Performance limits of matching pursuit algorithms,''
  \emph{International symposium on information theory}, pp. 2444--2448, 2008.

\bibitem{Tropp06}
J.~Tropp, ``Just relax: Convex programming methods for identifying sparse
  signal in noise,'' \emph{IEEE Trans. Info. Theo.}, vol.~52, no.~3, 2006.

\bibitem{Cover06}
T.~M. Cover and J.~A. Thomas, \emph{Elements of Information Theory}.\hskip 1em
  plus 0.5em minus 0.4em\relax Wiley, 2006.

\bibitem{Tse05}
D.~Tse and P.~Viswanath, \emph{Fundamentals of Wireless Communication}.\hskip
  1em plus 0.5em minus 0.4em\relax Cambridge University Press, 2005.

\bibitem{Hassibi00howmuch}
B.~Hassibi and B.~Hochwald, ``How much training is needed in multiple-antenna
  wireless links?'' \emph{IEEE Trans. Info. Theory}, vol.~49, pp. 951--963,
  2000.

\bibitem{Lapidoth96}
A.~Lapidoth, ``Nearest neighbor decoding for additive non-gaussian noise
  channels,'' \emph{IEEE Trans. Info. Theory}, vol.~42, no.~3, pp. 1520--1529,
  1996.

\bibitem{Resnick}
S.~I. Resnick, \emph{A probability path}.\hskip 1em plus 0.5em minus
  0.4em\relax Birkhauser Boston, 1999.

\bibitem{Rao05}
S.~F. Cotter, B.~D. Rao, K.~Engan, and K.~Kreutz-Delgado, ``Sparse solutions to
  linear inverse problems with multiple measurement vectors,'' \emph{IEEE
  Trans. Sig. Proc.}, vol.~53, no.~7, pp. 2477--2488, 2005.

\bibitem{Wipf_2007_b}
D.~P. Wipf and B.~D. Rao, ``An empirical bayesian strategy for solving the
  simultaneous sparse approximation problem,'' \emph{IEEE Trans. Signal
  Processing}, vol.~55, no.~7, pp. 3704--3716, July 2007.

\bibitem{JieChen}
J.~Chen and X.~Huo, ``Theoretical results on sparse representations of
  multiple-measurement vectors,'' \emph{IEEE Trans. Signal Proc.}, vol.~54,
  no.~12, pp. 4634--4643, Dec 2006.

\bibitem{Cichocki}
R.~Zdunek and A.~Cichocki, ``Improved m-focuss algorithm with overlapping
  blocks for locally smooth sparse signals,'' \emph{IEEE Trans. Signal Proc.},
  vol.~56, no.~10, pp. 4752--4761, Oct 2008.

\bibitem{Eldar_UnionSubspace}
Y.~C. Eldar and M.~Mishali, ``Robust recovery of signals from a structured
  union of subspaces,'' \emph{IEEE Transactions on Information Theory},
  vol.~55, no.~11, pp. 5302--5316.

\bibitem{Lancaster_x2}
H.~O. Lancaster, \emph{The $\chi^2$ distribution}.\hskip 1em plus 0.5em minus
  0.4em\relax Wiley, 1969.

\bibitem{CHENQI}
C.-P. Chen and F.~Qi, ``Completely monotonic function associated with the gamma
  functions and proof of wallis' inequality,'' \emph{Tamkang Journal of
  Mathematics}, vol.~36, pp. 303--307, 2005.

\bibitem{Silverstein}
J.~W. Silverstein, ``The smallest eigenvalue of a large dimensional wishart
  matrix,'' \emph{Ann. Probab.}, vol.~13, pp. 1364--1368, 1985.

\end{thebibliography}

\end{document}